\documentclass[12pt,a4paper]{article}
\usepackage{graphicx}
\usepackage[T1]{fontenc}
\usepackage[utf8]{inputenc}
\usepackage{textcomp}
\usepackage[sc,osf]{mathpazo}
\usepackage{a4wide}  
\usepackage{latexsym,amsthm,amsfonts,amsmath,mathrsfs,amssymb,amsbsy,amsgen}
\usepackage{accents}
\usepackage{enumerate}
\usepackage[nosort]{cite}
\usepackage{booktabs} 
\usepackage[unicode,implicit]{hyperref}
\hypersetup{%
  pdftitle    = {On the interactions and equilibrium between
      Einstein-Maxwell-Dilaton black holes}
  pdfkeywords = {Einstein-Maxwell-Dilaton, EMD, black holes, scalar fields,
    equilibrium, initial data problem, time-symmetric, Einstein-Rosen, bridge,
    hair, no-hair, interaction energy},
  pdfauthor   = {Ulrich K. Beckering Vinckers and Tom\'as Ort\'{\i}n},
  plainpages  = true,
  colorlinks  = true,
  citecolor   = blue,
  urlcolor    = red,
  linkcolor   = black
}
\newcommand{\hepth}[1]{{\tt
\href{http://www.arXiv.org/abs/hep-th/#1}{hep-th/#1}}}

\newcommand{\arxiv}[1]{{\tt arXiv:\href{http://www.arXiv.org/abs/#1}{#1}}}
\newcommand{\doi}[1]{{\tt DOI:\href{http://dx.doi.org/#1}{#1}}}

\allowdisplaybreaks

\makeatletter
\@addtoreset{equation}{section}
\makeatother

\begin{document}

\begin{flushright}
\small
IFT-UAM/CSIC-24-093\\
August 6\textsuperscript{th}, 2024\\
\normalsize
\end{flushright}

\vspace{1cm}

\begin{center}

  {\Large {\bf On the interactions and equilibrium between\\[.4cm]
      Einstein-Maxwell-Dilaton black holes}}

\vspace{1.5cm}

\renewcommand{\thefootnote}{\alph{footnote}}

{\sl\large Ulrich K.~Beckering Vinckers}$^{1,2,}$\footnote{Email: {\tt
  bckulr002[at]myuct.ac.za}}
{\sl\large and Tom\'as Ort\'{\i}n}$^{3,}$\footnote{Email: {\tt
    tomas.ortin[at]csic.es}}

\setcounter{footnote}{0}
\renewcommand{\thefootnote}{\arabic{footnote}}
\vspace{.8cm}

${}^{1}${\it Cosmology and Gravity Group, Department of Mathematics and
  Applied Mathematics, University of Cape Town, Rondebosch 7701, Cape Town,
  South Africa}\\[.3cm]

${}^{2}${\it Van Swinderen Institute, University of Groningen, 9747 AG
  Groningen, The Netherlands}\\[.3cm]

${}^{3}${\it Instituto de F\'{\i}sica Te\'orica UAM/CSIC
C/ Nicol\'as Cabrera, 13--15,  C.U.~Cantoblanco, E-28049 Madrid, Spain}

\vspace{.8cm}


{\bf Abstract}
\end{center}
\begin{quotation}
  {\small We study the interactions and the conditions for the equilibrium of
    forces between generic non-rotating black holes of the
    Einstein--Maxwell--Dilaton (EMD) theory. We study known (and some new)
    solutions of the time-symmetric initial-data problem describing an
    arbitrary number of those black holes, some of them with primary scalar
    hair. We show how one can distinguish between initial data corresponding
    to dynamical situations in which the black holes (one or many) are not in
    equilibrium and initial data which are just constant-time slices of a
    static solution of the full equations of motion describing static black
    holes using (self-)interaction energies. For a single black hole,
    non-vanishing self-interaction energy is always related to primary scalar
    hair and to a dynamical black hole. Removing the self-interaction energies
    in multi-center solutions we get interaction energies related to the
    attractive and repulsive forces acting on the black holes.  As shown by
    Brill and Lindquist, for widely separated black holes, these take the
    standard Newtonian and Coulombian forms plus an additional interaction
    term associated with the scalar charges which is attractive for like
    charges.}
\end{quotation}

\newpage
\pagestyle{plain}

\tableofcontents



\section{Introduction}

The study of the general-relativistic description of the gravitational
interaction between massive bodies and their motion beyond the idealized
test-particle limit is one of the most important research subjects in this
field, from the pioneering works of Einstein, Grommer, Infeld, Hoffman, and
Papapetrou among others
\cite{Einstein:1927,Einstein:1938yz,Fock:1939,Einstein:1940mt,Einstein:1949,Papapetrou:1951}
all the way to the recent detection of the gravitational radiation produced in
the gravitational interaction of two black holes
\cite{LIGOScientific:2016aoc}.

A very interesting approach to this problem is the study of particular
solutions to the initial-data problem that describe more than one gravitating
object (black holes, in general). This is a non-trivial problem which contains
a substantial deal of information about the interactions between the black
holes. In particular, as shown by Brill and Lindquist in
Refs.~\cite{Brill:1959zz,Brill:1963yv}, one can find the forces between the
back holes and the conditions for their equilibrium recovering, for large
separations, the Newtonian limit of the forces. Brill and Lindquist's approach
makes use of the great simplifications taking place in the so-called
time-symmetric case\footnote{See, \textit{e.g.}~Ref.~\cite{Gibbons:1972ym}.}
in which the Cauchy surface on which the initial-date problem is formulated
corresponds to a moment in which the black holes are instantaneously at rest,
but, usually, subject to non-vanishing accelerations.

In Ref.~\cite{Ortin:1995vg} Brill and Lindquist's study was extended to
theories including a scalar field. Some of the results obtained, corroborated
and further extended in Ref.~\cite{Cvetic:2014vsa}, were quite unusual, as we
are going to explain

In this approach, each black hole is associated with an Einstein--Rosen-like
bridge connecting two asymptotically-flat regions
\cite{Einstein:1935tc}. There is one asymptotically-flat region to which all
the bridges are connected and which corresponds to ``our universe'' in which
all the black holes are present at a given time and each bridge leads to
another universe in which only one black hole is present. One can compute the
total mass of the set of interacting black holes in our universe and the mass
of each individual black hole in each of the single-black-hole universes. The
total mass does not equal the sum of the individual masses and the difference
can be interpreted as the interaction energy between the black holes, from
which the force between two black holes can be extracted. For single black
holes of the Einstein--Maxwell theory considered by Brill and Lindquist in
Refs.~\cite{Brill:1959zz,Brill:1963yv}, the masses computed on both universes
are always identical and the interaction energies vanish, but in some of the
solutions found in Refs.~\cite{Ortin:1995vg,Cvetic:2014vsa} this was no longer
true: there are single black-hole initial data connecting two universes in
which the black hole seems to have different mass, scalar charge and
asymptotic value of the scalar. To the best of our knowledge, no physical
interpretation of this fact has been given so far.

In this paper we want to revisit this puzzle, put forward an explanation and
test it in different solutions of the time-symmetric initial-data problem of
the Einstein-Maxwell-Dilaton system, some of them, new.\footnote{The
  interactions between black-hole solutions of this system, specially, between
  the dyonic ones, have been recently studied in
  Ref.~\cite{Cremonini:2023vwf}.} We are going to argue that the difference in
the values of the masses and scalar charges in two universes is associated with
the non-staticity of the complete solution because, for static solutions, one
can construct a Komar charge
\cite{Komar:1958wp,Liberati:2015xcp,Ortin:2021ade,Mitsios:2021zrn} and a
scalar charge \cite{Pacilio:2018gom,Ballesteros:2023iqb} that satisfy
Gauss-type laws.\footnote{These charges are written as $(d-2)$-forms in $d$
  dimensions and one can show that they are on-shell closed if the spacetime
  admits a Killing vector. The closedness of the $(d-2)$-form charge is
  equivalent to the Gauss law. In our particular case, integrating the
  exterior derivative of the Komar charge (a 3-form that vanishes on shell)
  over the 3-dimensional Cauchy surface and applying Stokes theorem, we find
  that the sum of the integrals of the Komar charge 2-form over all the
  2-dimensional boundaries (2-spheres at infinity) must vanish. Each integral
  gives the mass in the corresponding asymptotic region and, taking into
  account the orientation of each boundary we find that the mass at each side
  of an Einstein--Rosen bridge for a single, static, black hole must take the
  same value.} This implies that the mass and scalar charge must necessarily
be the same in both sides of the Einstein--Rosen bridge when the total
spacetime is static. Then, the difference in masses and scalar charges implies
the non-staticity of the complete solution. The initial data correspond to a
dynamical black hole which is evolving and which, furthermore, has primary
scalar hair.\footnote{Secondary scalar hair, which is physically allowed
  \cite{Coleman:1991ku}, is related to the conserved charges (mass, electric
  and magnetic charges) by a fixed formula \cite{Ballesteros:2023iqb}. In all
  the cases that we have studied in which the masses and scalar charges are
  different in the two sides of the Einstein--Rosen bridge, the scalar charge
  is an independent parameter or combination of parameters and this relation
  is not satisfied. Thus, it corresponds to primary scalar hair.} The
evolution of a black hole which has primary scalar hair is a very interesting
problem related to the weak cosmic censorship and no-hair conjectures which
deserves further study but falls outside the scope of this work.

This paper is organized as follows: in Section~\ref{sec-thetheory} we present
the Einstein--Maxwell-Dilaton theory in our conventions. In
Section~\ref{sec-timesymmetricinitialdataproblem} we formulate the
time-symmetric initial data problem for this theory and in
Section~\ref{sec-knownexamples} we present and study different solutions of
increasing complexity: in Section~\ref{sec-schwarzschild} we review solutions
with vanishing scalar, electric and magnetic fields, showing how they describe
interacting Schwarzschild black holes. In Section~\ref{sec-JNW} we include a
non-trivial scalar field and we find solutions describing regular black holes,
with non-vanishing interaction energy even for the single black-hole case
(\textit{self-interaction energy}). When this self-interaction energy
vanishes, we show that the solutions are constant-time slices of the
(singular) Janis--Newman--Winicour solutions \cite{Janis:1968zz}. In
Section~\ref{sec-reissner-nordstrom} we consider non-trivial electric fields
but vanishing scalar (the Einstein--Maxwell theory), recovering the results
obtained by Brill and Lindquist in Ref.~\cite{Brill:1963yv}. In
Section~\ref{sec-magnetic-reissner-nordstrom} we describe how to generalize
the purely electric solutions of the previous section to purely magnetic or
dyonic ones. In Section~\ref{sec-EMD} we consider the solutions of
Ref.~\cite{Cvetic:2014vsa}, which have non-trivial electric and scalar
fields. We show how, when the self-interaction energies vanish, the solutions
describe constant-time slices of the well-known static, regular, electric
dilaton black-hole solutions of
Refs.~\cite{Gibbons:1982ih,Gibbons:1984kp,Holzhey:1991bx} which have no
primary scalar hair. Then, eliminating these self-interaction energies, we can
determine the interaction energy and force between two of these black holes,
recovering the expected result.  We generalize the results of this section to
the dyonic $a=1$ case in Section~\ref{sec-magnetic-EMD}. We discuss our
results and possible future directions in
Section~\ref{sec-discussion}. Finally, the appendices contain some of the
complete spacetime solutions mentioned in the main text and their
transformation to coordinates that allow us to get constant-time slices at the
moment of time-symmetry showing the two sides of the Einstein--Rosen bridge.

\section{The theory}
\label{sec-thetheory}

The family of theories that we are going to consider are collectively known in
the literature as the Einstein-Maxwell-Dilaton (EMD) model. The action, in
differential-form language, is given by\footnote{We set $G_{N}^{(4)}=1$ in
  this paper.}

\begin{equation}
  \label{eq:actiondilatonBHs}
    S[e,A,\phi]
     =
      \frac{1}{16\pi}
      \int \left\{ -\star(e^{a}\wedge e^{b})
      \wedge R_{ab}
      +\tfrac{1}{2}d\phi\wedge \star d\phi
      +\tfrac{1}{2}e^{-a\phi}F\wedge \star F \right\}\,,
\end{equation}

\noindent
and contains the real parameter $a$ which determines the strength of the
coupling of the dilaton $\phi$ (a real scalar field) to the Maxwell field
$A=A_{\mu}dx^{\mu}$ with field strength $F=dA$. Different values of $a$
correspond to different theories. 

The (left-hand sides of the) equations of motion are defined by the general
variation of the action generated by an arbitrary variation of the fields
(here $\varphi$ stands for all the fields of the theory)

\begin{equation}
  \delta S
  =
  \int\left\{
    \mathbf{E}_{a}\wedge \delta e^{a} + \mathbf{E}_{\phi}\delta\phi
    +\mathbf{E}_{A}\wedge \delta A
    +d\mathbf{\Theta}(\varphi,\delta\varphi)
    \right\}\,,
\end{equation}

\noindent
and are given by

\begin{subequations}
  \begin{align}
    \label{eq:Ea}
    \mathbf{E}_{a}
    & =
      \imath_{a}\star(e^{b}\wedge e^{c})\wedge R_{bc}
      +\tfrac{1}{2}\left(\imath_{a}d\phi \star d\phi
      +d\phi\wedge \imath_{a}\star d\phi\right)
      \nonumber \\
    & \nonumber \\
    & \hspace{.5cm}
      -\frac{1}{2}e^{-a\phi}\left(\imath_{a}F\wedge\star F
      -F\wedge \imath_{a}\star F\right)\,,
    \\
    & \nonumber \\
    \mathbf{E}_{\phi}
    & =
      -\left(d\star d\phi
      +\frac{a}{2}e^{-a\phi} F\wedge\star F\right)\,,
    \\
    & \nonumber \\
    \mathbf{E}_{A}
    & =
      -d\left( e^{-a\phi}\star F\right)\,.
  \end{align}
\end{subequations}

Since we are going to work with electric and magnetic fields, we have to add
to these the Bianchi identity

\begin{equation}
  \mathbf{B}_{A}
  =
  d F\,.
\end{equation}

\section{The time-symmetric initial-data problem}
\label{sec-timesymmetricinitialdataproblem}

Let $\hat{n}=n_{\mu}dx^{\mu}$ be the 1-form dual to the unit-norm, timelike,
vector field $n=n^{\mu}\partial_{\mu}$ normal to the Cauchy surface $\Sigma$,
$\imath_{n}\hat{n}=n^{\mu}n_{\mu}=1$. Taking into account our mostly minus
signature, the metric induced on the Cauchy hypersurface is given by

\begin{equation}
h = g-\hat{n}\otimes \hat{n}\,,  
\end{equation}

\noindent
or, in components,

\begin{equation}
h_{\mu\nu} = g_{\mu\nu}-n_{\mu}n_{\nu}\,.    
\end{equation}

In order to project over the 3-dimensional surface we can use

\begin{equation}
1^{(3)} = 1 - \hat{n}\imath_{n}\,, 
\end{equation}

\noindent
which is equivalent to

\begin{equation}
h_{\mu}{}^{\nu} = g_{\mu}{}^{\nu}-n_{\mu}n^{\nu}\,.    
\end{equation}

The extrinsic curvature is defined over the Cauchy surface as

\begin{equation}
  K
  =
  \tfrac{1}{2}\pounds_{n}h\,.
\end{equation}

In the time-symmetric case
\cite{Misner:1957mt,Brill:1959zz,Brill:1963yv,Gibbons:1972ym} the normal unit
vector is covariantly constant on $\Sigma$

\begin{equation}
  \nabla n
    \stackrel{\Sigma}{=}
0\,,
\end{equation}

\noindent
which implies 

\begin{equation}
d\hat{n}
\stackrel{\Sigma}{=}
0\,,
\,\,\,\,
\Rightarrow
\,\,\,\,
\hat{n}\wedge d\hat{n}
\stackrel{\Sigma}{=}
0\,,
\end{equation}

\noindent
so it is hypersurface orthogonal.  All this means that, if we define the time
coordinate by

\begin{equation}
n \equiv \partial_{t}\,,  
\end{equation}

\noindent
with $\Sigma$ defined by $t=0$, the metric induced on $\Sigma$ is
time-independent and of the form

\begin{equation}
  ds^{2}
  \stackrel{\Sigma}{=}
  h_{mn}dx^{m}dx^{n}\,,
  \,\,\,\,\,
  m,n=1,2,3\,.
\end{equation}

\noindent
In the time-symmetric case we also have

\begin{equation}
  \imath_{n}d\phi
  \stackrel{\Sigma}{=}
  0\,.
\end{equation}

In the time-symmetric initial data problem we only need to consider the
projection of the equations of motion with $n$, taking into account that the
extrinsic curvature, $K$, vanishes over the surface of time symmetry because
$n$ is covariantly constant there.  The equations to consider are
$\hat{n} \wedge n^{a}\mathbf{E}_{a}$, $\hat{n}\wedge \mathbf{E}_{A}$ and
$\hat{n}\wedge \mathbf{B}_{A}$ evaluated over $\Sigma$. It is convenient to
define the electric and magnetic 1-forms relative to $n$ as

\begin{equation}
  E
  \equiv
  \imath_{n}F\,,
  \hspace{1cm}
  B
  \equiv
  e^{-a\phi}\imath_{n}\star F\,.
\end{equation}

The field strength and its Hodge dual can be written in terms of these 1-forms
as follows:

\begin{subequations}
  \begin{align}
  F
  & =
    \hat{n}\wedge E -e^{a\phi}\star(\hat{n}\wedge B)\,,
    \\
    & \nonumber \\
  \hspace{1cm}
  \star F
  & =
  e^{a\phi}\hat{n}\wedge B+\star (\hat{n}\wedge E)\,.
  \end{align}
\end{subequations}

Let us consider first the projection of the Einstein equations:

\begin{equation}
  \begin{aligned}
        \hat{n}\wedge n^{a}\mathbf{E}_{a}
    & \stackrel{\Sigma}{=}
      \hat{n}\wedge \imath_{n}\star(e^{b}\wedge e^{c})\wedge R_{bc}
      -\tfrac{1}{2} \hat{n}\wedge d\phi\wedge\star (\hat{n}\wedge d\phi)
      \\
      & \\
      & \hspace{.5cm}
      -\tfrac{1}{2}e^{-a\phi}
      ( \hat{n}\wedge E) \wedge \star (\hat{n}\wedge E)
        -\tfrac{1}{2}e^{a\phi}
        (\hat{n}\wedge B)\wedge \star(\hat{n}\wedge B)\,.
  \end{aligned}
\end{equation}

Next, let us consider the projection of the Maxwell equation

\begin{equation}
  \begin{aligned}
        \hat{n}\wedge \mathbf{E}_{A}
    & \stackrel{\Sigma}{=}
   \hat{n}\wedge  d\left[e^{-a\phi}\star^{(3)} E \right]\,,
  \end{aligned}
\end{equation}

\noindent
where $\star^{(3)}$ is the Hodge operator on $\Sigma$ with the induced metric.

Following the same steps,  the Bianchi identity can be brought to the form

\begin{equation}
  \begin{aligned}
    \hat{n}\wedge \mathbf{B}_{A}
    & \stackrel{\Sigma}{=}
-\hat{n}\wedge  d\left[e^{a\phi}\star^{(3)} B \right]\,.
  \end{aligned}
\end{equation}

These two equations are related by electric-magnetic duality transformations
that interchange $E$ and $B$ and reverse the sign of $\phi$:

\begin{equation}
  \label{eq:electric-magnetic}
  E'=B\,,
  \hspace{1cm}
  B'= -E\,,
  \hspace{1cm}
  \phi'=-\phi\,.
\end{equation}

The projection of the Einstein equations is also invariant under them.

It is now convenient to use a Vielbein basis $\{e^{a}=e^{a}{}_{\mu}dx^{\mu}\}$
in which $e^{0}=\hat{n}$. The rest of the 1-forms will be denoted by $e^{i}$,
$i=1,2,3$ and, on $\Sigma$, they are a Dreibein for the 3-dimensional metric
$h_{mn}$:\footnote{A more precise (but also more involved) notation would
  distinguish between the elements of the Vierbein $e^{i}$ and their pullbacks
  over $\Sigma$, which constitute a Dreibein of the induced metric. We will
  assume from this moment that all the objects and equations are 3-dimensional
  and defined on $\Sigma$ only.}

\begin{equation}
  ds^{2}
  \stackrel{\Sigma}{=}
  h_{mn}dx^{m}dx^{n}
  \stackrel{\Sigma}{=}  
  -\delta_{ij}e^{i}e^{j}\,,
  \hspace{1cm}
  E= E_{i}e^{i}\,,
  \hspace{1cm}
  B= B_{i}e^{i}\,,
\end{equation}

\noindent
In this basis, the above equations take the form

\begin{subequations}
  \begin{align}
    R(-h) -\tfrac{1}{2}\left[\partial_{i}\phi\partial_{i}\phi
+e^{-a\phi}E_{i}E_{i}+e^{+a\phi}B_{i}B_{i}
    \right]
    & =
      0\,,    \\
    & \nonumber \\
    \mathcal{D}_{i}\left(e^{-a\phi}E_{i}\right)
    & =
      0\,,
    \\
    & \nonumber \\
    \mathcal{D}_{i}\left(e^{a\phi}B_{i}\right)
    & =
      0\,,
  \end{align}
\end{subequations}

\noindent
where $R(-h)$ is the Ricci scalar of the 3-dimensional metric with positive
signature.

Our ansatz for the 3-dimensional metric is always going to be of the form

\begin{equation}
  -h_{mn}dx^{m}dx^{n}
  =
  \Phi^{4}\bar{h}_{mn}dx^{m}dx^{n}\,,
\end{equation}

\noindent
where $\bar{h}_{mn}$ is a reference, background, 3-dimensional metric of
positive signature, such as the Euclidean metric, the metric of the round
S$^{3}$ or that of the product S$^{2}\times$S$^{1}$ \cite{Cvetic:2014vsa}. For
this ansatz,

\begin{equation}
  R(-h)
  =
   -\Phi^{-4}\left[\bar{R}+8\Phi^{-1}\bar{\nabla}^{2}\Phi\right]\,,
\end{equation}

\noindent
where the barred objects are computed with the barred metric. Then, the above
three equations can be written in the background metric in the form

\begin{subequations}
  \begin{align}
  \bar{R}+8\Phi^{-1}\bar{\nabla}^{2}\Phi
  +\tfrac{1}{2}\bar{h}^{mn}\left[\partial_{m}\phi\partial_{n}\phi
+e^{-a\phi}E_{m}E_{n}+e^{a\phi}B_{m}B_{n}
    \right]
    & =
      0\,,    \\
    & \nonumber \\
    \bar{\nabla}_{m}\left(\Phi^{2}e^{-a\phi}E^{m}\right)
    & =
      0\,,
    \\
    & \nonumber \\
    \bar{\nabla}_{m}\left(\Phi^{2}e^{a\phi}B^{m}\right)
    & =
      0\,.
  \end{align}
\end{subequations}

Finally, if the background metric is the 3-dimensional Euclidean metric
$\mathbb{E}^{3}$ in Cartesian coordinates, the above equations take the simple
form

\begin{subequations}
  \label{eq:finalequations}
  \begin{align}
    \label{eq:Einstein}
  8\Phi^{-1}\partial_{m}\partial_{m}\Phi
  +\tfrac{1}{2}\left[\partial_{m}\phi\partial_{m}\phi
+e^{-a\phi}E_{m}E_{m}+e^{a\phi}B_{m}B_{m}
    \right]
    & =
      0\,,    \\
    & \nonumber \\
    \label{eq:Gauss1}
    \partial_{m}\left(\Phi^{2}e^{-a\phi}E_{m}\right)
    & =
      0\,,
    \\
    & \nonumber \\
    \label{eq:Gauss2}
    \partial_{m}\left(\Phi^{2}e^{a\phi}B_{m}\right)
    & =
      0\,.
  \end{align}
\end{subequations}

This is the case that we are going to consider in this paper and these are the
PDEs for which we will find solutions describing interacting black holes at
the moment of time symmetry. 

\section{Solutions to the time-symmetric initial-data problem}
\label{sec-knownexamples}

In this section we are going to reproduce and interpret physically some
well-known and some new solutions of the time-symmetric data problem equations
for the EMD model, Eqs.~(\ref{eq:finalequations}). We will first consider the
vacuum case and later we will consider the addition of matter fields of the
EMD theory, gradually increasing the complexity of the solutions and
introducing the concepts that we need. Most of the solutions and their generic
interpretation as interacting black holes are not new. However, most of the
solutions which have non-trivial scalar fields exhibit strange features (see
Ref.~\cite{Ortin:1995vg}) for which we are going to propose a physical
interpretation related to the static or dynamical nature of the complete
spacetime solution.  This will allow us to determine the forces acting between
two widely-separated EMD black holes.

\subsection{Schwarzschild black holes}
\label{sec-schwarzschild}

Let us first consider solutions with vanishing electric and magnetic fields
and constant scalar field. The only equation left to be solved is the Laplace
equation in $\mathbb{E}^{3}$

\begin{equation}
  \partial_{m}\partial_{m}\Phi
  =
  0\,.
\end{equation}

\noindent
A convenient set of asymptotically-flat solutions is provided by

\begin{equation}
  \Phi
  =
  1 +\sum_{i=1}^{N}\frac{\Phi_{i}/2}{|\vec{x}-\vec{x}_{i}|}\,,
\end{equation}

\noindent
where the vectors $\vec{x}_{i}$ are constant.

For $N=1$, setting $\vec{x}_{1}=\vec{0}$ and using spherical coordinates

\begin{equation}
  d\vec{x}^{\,2}
  =
  d\rho^{2}+\rho^{2}d\Omega_{(2)}^{2}\,,
  \,\,\,\,\,
  \text{with}
  \,\,\,\,\,
  \rho^{2} = |\vec{x}|^{2}
  \,\,\,\,\,
  \text{and}
  \,\,\,\,\,
  d\Omega_{(2)}^{2}
  =
  d\theta^{2}+\sin^{2}{\theta}d\varphi^{2}\,,
\end{equation}

\noindent
we recover the spatial part of the Schwarzschild metric with mass $M=\Phi_{1}$
in isotropic coordinates, given in Eq.~(\ref{eq:Schwarzschildisotropic}) which
has the event horizon located at $\rho=M/2$.

In the $\rho\rightarrow 0$ limit, the 3-dimensional metric 

\begin{equation}
\left(1+\frac{M/2}{\rho}\right)^{4}\left(d\rho^{2}+\rho^{2}d\Omega_{(2)}^{2}\right)\,,  
\end{equation}

\noindent
becomes singular, but it can be analytically continued using
$\rho' \equiv (M/2)^{2}/\rho$. The metric takes exactly the same form, with
$\rho$ replaced by $\rho'$ and now covers another asymptotically-flat region,
which corresponds to a constant-time slice of region III of the
Kruskal-Szekeres spacetime. We conclude that these two charts cover the $T=0$
(time-symmetric) slice of the Kruskal-Szekeres spacetime, which is also known
as the \textit{Einstein-Rosen bridge} \cite{Einstein:1935tc} of a black hole
of mass $M$. The ``neck'' of the bridge coincides with the intersection of the
$T=0$ slice with the event horizon, which is the bifurcation sphere.

When $N>1$ we can use spherical coordinates centered at the
$i^{th}$ pole of the harmonic function $\Phi$

\begin{equation}
  d(\vec{x}-\vec{x}_{i})^{\,2}
  =
  d\rho_{i}^{2}+\rho_{i}^{2}d\Omega_{(2)i}^{2}\,,
  \,\,\,\,\,
  \text{with}
  \,\,\,\,\,
  \rho_{i}^{2} = |\vec{x}-\vec{x}_{i}|^{2}
  \,\,\,\,\,
  \text{and}
  \,\,\,\,\,
  d\Omega_{(2)i}^{2}
  =
  d\theta_{i}^{2}+\sin^{2}{\theta_{i}}d\varphi_{i}^{2}\,,
\end{equation}

\noindent
and analytically continue the metric in the $\rho_{i}\rightarrow 0$ direction
using a new coordinate $\rho_{i}' = (\Phi_{i}/2)^{2}/\rho_{i}$. First, in the
$\rho_{i}\to 0$ limit

\begin{equation}
  \begin{aligned}
    \Phi
    & =
    \frac{\Phi_{i}/2}{\rho_{i}}+ A_{i}
    +\mathcal{O}(\rho_{i})
  \end{aligned}
\end{equation}

\noindent
where

\begin{equation}
  A_{i}
  \equiv
  1+\sum_{j\neq i}\frac{\Phi_{j}/2}{|\vec{x}_{ij}|}\,,
  \hspace{1cm}
  \vec{x}_{ij}
  \equiv
  \vec{x}_{j}-\vec{x}_{i}\,,
\end{equation}

\noindent
and there is no sum over the repeated indices.  In the new coordinates

\begin{equation}
  \Phi
  =
  \frac{\rho_{i}'}{\Phi_{i}/2}
  \left[\left(1+ \frac{A_{i}\Phi_{i}/2}{\rho_{i}'}\right)
    +\mathcal{O}(1/\rho_{i}^{\prime\, 2})\right]\,, 
\end{equation}

\noindent
and the full 3-dimensional metric is

\begin{equation}
\left[1+ \frac{2A_{i}\Phi_{i}}{\rho_{i}'}
  +\mathcal{O}(1/\rho_{i}^{\prime\, 2})\right]
\left(d\rho_{i}^{\prime\, 2}+\rho_{i}^{\prime\, 2}d\Omega_{(2)i}^{2}\right)\,.
\end{equation}

Comparing with the single black-hole case, we see that, in the $\rho_{i}\to 0$
limit, this metric approaches that of a constant time slice of the metric of a
single Schwarzschild black hole of mass

\begin{equation}
  M_{i}
  =
  \Phi_{i}A_{i}
  =
  \Phi_{i}
  +\Phi_{i}\sum_{j\neq i}\frac{\Phi_{j}/2}{|\vec{x}_{ij}|}\,.
\end{equation}

This implies that the original Cartesian coordinates cover a patch of the
Cauchy hypersurface with $N$ interacting Schwarzschild black holes. We can
only compute their total mass in that asymptotically-flat region. Expanding
the metric for large values of the three Cartesian coordinates and defining
$\rho\equiv |\vec{x}|$, the 3-dimensional metric behaves as

\begin{equation}
\sim \left[1
  +\frac{2\sum_{i=1}^{N}\Phi_{i}}{\rho}+\mathcal{O}(1/\rho^{2})\right]
\left(d\rho^{2}+\rho^{2}d\Omega_{(2)}^{2}\right)\,,
\end{equation}

\noindent
and the total mass is

\begin{equation}
  M
  =
  \sum_{i=1}^{N}\Phi_{i}\,.
\end{equation}

The \textit{interaction energy} can be defined as the difference between the
the total mass and the sum of the individual masses \cite{Brill:1963yv}

\begin{equation}
  M_{int}
  \equiv
  M-\sum_{i=1}^{N}M_{i} 
  =
  -\sum_{i}^{N}\Phi_{i}\sum_{j\neq i}\frac{\Phi_{j}/2}{|\vec{x}_{ij}|}
  =
  -\sum_{i<j}\frac{\Phi_{i}\Phi_{j}}{|\vec{x}_{ij}|}\,.
\end{equation}

For $N=2$ we can express the integration constants $\Phi_{1,2}$ in terms of
the individual masses $M_{1,2}$ and the ``distance''\footnote{$|\vec{x}_{12}|$
  would be the distance if the metric was the Euclidean metric, which is only
  the case asymptotically.} $|\vec{x}_{12}|$ exactly, but the result is not
very illuminating. For large separations,

\begin{equation}
  \Phi_{1,2}
  =
  M_{1,2}- \frac{M_{1}M_{2}/2}{|\vec{x}_{12}|}+\mathcal{O}(1/|\vec{x}_{12}|^{2})\,,
\end{equation}

\noindent
and the interaction energy

\begin{equation}
  M_{int}
  =
  -\frac{M_{1}M_{2}}{|\vec{x}_{12}|}\left[1 -
    \frac{(M_{1}+M_{2})/2}{|\vec{x}_{12}|}
  +\mathcal{O}\left(\frac{1}{|\vec{x}_{12}|^{2}}\right)\right]\,,
\end{equation}

\noindent
approaches the Newtonian value when $|\vec{x}_{12}|\rightarrow \infty$.

Although in this case this observation looks quite trivial, we would like to
stress the fact that, for the single Schwarzschild black hole, the mass can be
computed in both coordinate patches (that is, in both asymptotic regions) and
the values obtained are identical, which we can trivially and obviously
associate to the vanishing of the interaction energy for a single black
hole. However, in some of the cases that we are going to consider next, the
interaction energy for a single black hole (which we are going to call
\textit{self-interaction energy}) will not vanish: the masses computed in the
two asymptotically-flat regions at the ends of the Einstein--Rosen bridge will
be different, a fact that demands a physical interpretation.

At this point it is convenient to remember that, for each Killing vector of a
given spacetime, the vacuum Einstein equations admit an on-shell closed
(``conserved'') 2-form charge which is the Noether charge associated with the
invariance of the theory under general coordinate transformations: the Komar
charge \cite{Komar:1958wp}. In stationary spacetimes, when the Killing vector
is the one that generates time translations, the integral of this 2-form at
spatial infinity gives the ADM mass. Since the Komar 2-form is closed, it
satisfies a Gauss law and, if there are two asymptotic regions whose union is
the boundary of a Cauchy surface, the integrals at both spatial infinities
must shed the same value for the ADM masses (after the orientation is taken
into account). Thus, the identity of the ADM masses in the two asymptotic
regions is associated with the staticity of the complete solution and, in the
case at hand, this fact looks trivial because we knew beforehand that the
solution of the initial-data problem is just a constant-time slice of the
Schwarzschild solution. This identity of the masses, however, can be used as a
test of the stationarity of the full solution when we only know the initial
data and single black hole initial data which have different masses in
different asymptotic regions must correspond to non-stationary (``dynamical'')
solutions of the complete equations of motion in which the black hole will
evolve towards a different, stationary solution.

It should also be mentioned that it is possible to define scalar charges that
only satisfy a Gauss law in stationary spacetimes
\cite{Pacilio:2018gom,Ballesteros:2023iqb}. In stationary,
spherically-symmetric spacetimes, these charges are essentially the
coefficients of the $1/r$ terms in the asymptotic expansions of the scalar
fields in standard spherical coordinates. When these charges are different in
the two asymptotic regions of the initial data of a single black hole, as in
the case studied in Ref.~\cite{Ortin:1995vg}, the complete spacetime cannot
be stationary. Again, we expect the initial data to evolve towards a
stationary solution.

The evolution of the initial data sets with different masses and scalar
charges in different asymptotic regions may provide interesting insights on
the no-hair conjecture (will primary scalar hair be present in the final
state?)  and on cosmic censorship (will the final state exhibit naked
singularities?), but it has to be studied by other means which fall out of the
scope of this paper.

Let us move to the next case, in which the phenomenon observed in
Ref.~\cite{Ortin:1995vg} appears for the first time.

\subsection{Janis--Newman--Winicour solutions}
\label{sec-JNW}

Let us now consider a non-vanishing scalar field but vanishing electric and
magnetic fields.  Following Ref.~\cite{Ortin:1995vg}, we make the following
ansatz for the metric function $\Phi$ and the scalar field $\phi$ in terms of
two functions $\chi$ and $\psi$ which we want to be harmonic in 3-dimensional
Euclidean space:

\begin{subequations}
  \begin{align}
    \Phi^{4}
    & =
      \psi^{\delta} \chi^{\gamma}\,,
    \\
    & \nonumber \\
    \phi
    & =
      \phi_{0}+\alpha\log{\psi}+\beta\log{\chi}\,.
  \end{align}
\end{subequations}

Eq.~(\ref{eq:Einstein}) is solved by \cite{Ortin:1995vg}

\begin{subequations}
  \begin{align}
    \Phi^{4}
    & =
      \psi^{2} \chi^{2}(\psi/\chi)^{2\alpha}\,,
    \\
    & \nonumber \\
    \phi
    & =
      \phi_{0}\pm 2\sqrt{1-\alpha^{2}}\log{(\psi/\chi)}\,,
    \\
    & \nonumber \\
    \partial_{m}\partial_{m}\psi
    & =
    \partial_{m}\partial_{m}\chi
      =
      0\,,
  \end{align}
\end{subequations}

\noindent
where the constant $\alpha$ satisfies $|\alpha|\leq 1$ and can be taken to be
positive for simplicity. A convenient choice for the harmonic functions
$\psi,\chi$ which gives an asymptotically-flat metric is

\begin{subequations}
  \begin{align}
    \psi
    & =
      \psi_{0}\left(1 +\sum_{i=1}^{N}\frac{\psi_{i}}{|\vec{x}-\vec{x}_{i}|}\right)\,,
    \\
    & \nonumber \\
    \chi
    & =
      \chi_{0}\left(1 +\sum_{i=1}^{N}\frac{\chi_{i}}{|\vec{x}-\vec{x}_{i}|}\right)\,,
  \end{align}
\end{subequations}

\noindent
where the constants $\psi_{0},\psi_{i},\chi_{0},\chi_{i}$ are also taken to be
positive to avoid singularities.

The normalization of the metric at infinity in the patch covered by the
Cartesian coordinates demands

\begin{equation}
  \chi_{0}
  =
  \psi_{0}^{\frac{\alpha+1}{\alpha-1}}\,.
\end{equation}

\noindent
This condition makes them disappear from the metric and we can also make them
disappear from the dilaton through a redefinition of the constant
$\phi_{0}$. Hence, we will set them to $1$ in what follows.

In the same patch, we can compute the asymptotic value of the dilaton
$\phi_{\infty}$, the total mass $M$ and the total scalar charge $\Sigma$
(simply defined as the coefficient of $1/\rho$ in the asymptotic expansion of
the dilaton\cite{Pacilio:2018gom,Ballesteros:2023iqb}):

\begin{subequations}
  \begin{align}
    \phi_{\infty}
    & =
      \phi_{0}\,,
    \\
    & \nonumber \\
    M
    & =
      (1+\alpha)\sum_{i=1}^{N}\psi_{i}
      + (1-\alpha)\sum_{i=1}^{N}\chi_{i}\,, 
    \\
    & \nonumber \\
    \Sigma
    & =
      \pm 2\sqrt{1-\alpha^{2}}
      \sum_{i=1}^{N}\left(\psi_{i}-\chi_{i}\right)\,.
  \end{align}
\end{subequations}

Let us now consider the single center solution $N=1$, setting
$\vec{x}_{1}=\vec{0}$ and defining $\rho \equiv \vec{x}$.

It is easy to see that, when $\psi_{1}=-\chi_{1}=\omega/4$, this solution is a
constant-time slice of the static Janis--Newman--Winicour (JNW) solution
\cite{Janis:1968zz} in isotropic coordinates, given in
Eqs.~(\ref{eq:JNWisotropic}), which is singular for any $\Sigma\neq 0$ and
becomes Schwarzschild's for $\Sigma=0$. For positive values of the integration
constants $\chi_{1},\psi_{1}$ we get more general solutions which look
singular in the $\rho \rightarrow 0$ limit but can be analytically extended to
another asymptotically-flat region using the new coordinate

\begin{equation}
\frac{\rho}{\psi_{1}^{(1+\alpha)/2}\chi_{1}^{(1-\alpha)/2}}
=
\frac{\psi_{1}^{(1+\alpha)/2}\chi_{1}^{(1-\alpha)/2}}{\rho'}\,.
\end{equation}

In this new region, the metric and the dilaton take the form

\begin{subequations}
  \begin{align}
    \Phi^{3}d\vec{x}^{\, 2}
    & =
  \left(1+\frac{\psi_{1}^{\alpha}\chi_{1}^{1-\alpha}}{\rho'}\right)^{2(1+\alpha)}
  \left(1+\frac{\psi_{1}^{1+\alpha}\chi_{1}^{-\alpha}}{\rho'}\right)^{2(1-\alpha)}
      \left(d\rho^{\prime\, 2}+\rho^{\prime\, 2}d\Omega_{(2)}^{2}\right)\,,
    \\
    & \nonumber \\
    \phi
    & =
      \phi_{0}\pm 2\sqrt{1-\alpha^{2}}\log{\frac{\psi_{1}}{\chi_{1}}}
       \pm 2\sqrt{1-\alpha^{2}}\log{\frac{ \left(1+\frac{\psi_{1}^{\alpha}\chi_{1}^{1-\alpha}}{\rho'}\right)}{\left(1+\frac{\psi_{1}^{1+\alpha}\chi_{1}^{-\alpha}}{\rho'}\right)}}\,.
  \end{align}
\end{subequations}

\noindent
Thus, for $\chi_{1},\psi_{1}>0$, this solution describes an
Einstein-Rosen-type bridge potentially associated with a black hole.

The asymptotic values of the dilaton, mass and scalar charge measured in this
second asymptotic region are

\begin{subequations}
  \begin{align}
    \phi_{\infty}'
    & =
      \phi_{0}\pm 2\sqrt{1-\alpha^{2}}\log{\frac{\psi_{1}}{\chi_{1}}}\,,
    \\
    & \nonumber \\
    M'
    & =
      (1+\alpha)\psi_{1}^{\alpha}\chi_{1}^{1-\alpha}
      + (1-\alpha)\psi_{1}^{1+\alpha}\chi_{1}^{-\alpha}\,, 
    \\
    & \nonumber \\
    \Sigma
    & =
      \pm 2\sqrt{1-\alpha^{2}}
      \left(\psi_{1}^{\alpha}\chi_{1}^{1-\alpha}-\psi_{1}^{1+\alpha}\chi_{1}^{-\alpha}\right)\,,
  \end{align}
\end{subequations}

\noindent 
and are, generically, different from those computed in the first asymptotic
region \cite{Ortin:1995vg}.

According to the discussion at the end of the preceding section, this
difference is due to the non-stationary nature of the full spacetime
corresponding to these initial data. Therefore, these initial data describe
black holes which will evolve, settling to the only known static solutions of
this theory: the JNW solutions which include the Schwarzschild black hole for
vanishing scalar charge. In its evolution the black hole may lose all its
scalar charge and end up as a Schwarzschild black hole, with a singularity
hidden under an event horizon, or may fail to do it, ending up as a singular
JNW solution, violating the weak cosmic censorship conjecture. This is clearly
an important issue that deserves further investigation.

Finally, observe that, for $\alpha=0$, with $\chi_{1}=0$ (the case
$\psi_{1}=0$ is equivalent), the $\rho\rightarrow 0$ limit of the solution can
be brought to the form

\begin{subequations}
  \begin{align}
    \label{eq:cylindrical}
  \Phi^{4}d\vec{x}^{\, 2}
  & \sim
    \psi_{1}^{2} dx^{2}+\psi_{1}^{2}d\Omega_{(2)}^{2}\,,
    \\
    & \nonumber \\
    \phi
    & \sim
      \mp 2 x\,, 
  \end{align}
\end{subequations}

\noindent
which describes an infinite cylindrical throat with spherical (S$^{2}$)
section of radius $\psi_{1}$ and with a linear dilaton. In this region we
cannot compute the mass in the standard fashion. The infinite, cylindrical
throat is characteristic of extremal, static black holes and its occurrence in
a solution of this kind is new. It is unclear to which solution of the full
4-dimensional system it corresponds to.

\subsection{Electric Reissner--Nordstr\"om  black holes}
\label{sec-reissner-nordstrom}

Next, we are going to consider charged solutions of the Einstein--Maxwell (EM)
theory, which correspond to the $a=0$ case of Eq.~(\ref{eq:actiondilatonBHs})
with constant scalar (which we are going to ignore).

In this case, following Ref.~\cite{Brill:1963yv}, we make an ansatz based on 2
would-be harmonic functions:

\begin{subequations}
  \begin{align}
    \Phi^{4}
    & =
      \sigma^{\delta}  \kappa^{\gamma}\,,
    \\
    & \nonumber \\
    E_{m}
    & =
      \partial_{m}\left(\alpha\log{\sigma}+\beta\log{\kappa}\right)\,.
  \end{align}
\end{subequations}

This ansatz leads exactly to the same form of Eq.~(\ref{eq:Einstein}) as the
ansatz for the JNW solutions and, therefore, admits the same solutions. It is
not difficult to see that Eq.~(\ref{eq:Gauss1}) forces the parameter that we
called $\alpha$ in the previous case to vanish. Thus, we arrive at the
following solutions of the initial data problem \cite{Brill:1963yv}

\begin{subequations}
  \begin{align}
    \Phi^{4}
    & =
      \sigma^{2} \kappa^{2}\,,
    \\
    & \nonumber \\
    E_{m}
    & =
      2\partial_{m}\log{(\sigma/\kappa)}\,,
    \\
    & \nonumber \\
    \partial_{m}\partial_{m}\sigma
    & =
      \partial_{m}\partial_{m}\kappa
      =
      0\,.
  \end{align}
\end{subequations}

The electrostatic potential $P$ is defined by 

\begin{equation}
  E_{m}
  =
  \partial_{m}P\,,
\end{equation}

\noindent
up to an additive constant that should be physically irrelevant. We can take

\begin{equation}
  \label{eq:electrostaticpotentialRNcase}
  P
  =
 2\log{(\sigma/\kappa)}\,.
\end{equation}

Again, a convenient choice for the harmonic functions $\sigma,\kappa$ that
gives an asymptotically-flat metric with the standard normalization is

\begin{subequations}
  \begin{align}
    \sigma
    & =
      1 +\sum_{i=1}^{N}\frac{\sigma_{i}}{|\vec{x}-\vec{x}_{i}|}\,,
    \\
    & \nonumber \\
    \kappa
    & =
      1 +\sum_{i=1}^{N}\frac{\kappa_{i}}{|\vec{x}-\vec{x}_{i}|}\,,
  \end{align}
\end{subequations}

\noindent
where the integration constants $\sigma_{i}$ and $\kappa_{i}$ have to be
positive to avoid singularities. It is not too difficult to see that, for a
single center ($N=1$) these solutions are just constant-time slices of the
Reissner--Nordstr\"om (RN) solutions given in
Eqs.~(\ref{eq:electricRNsolutionsisotropic}) and
(\ref{eq:electricRNsolutionsisotropicintegrationconstants}).\footnote{At first
  sight, the RN solutions depend on three functions. However, the identity
  $\alpha (2\psi_{1}-\sigma_{1}-\kappa_{1}) = 2(\sigma_{1}-\kappa_{1})$ allows
  us to rewrite the constant-time slices (in particular, the electric field)
  entirely in terms of just $\sigma$ and $\kappa$. Observe that
  \begin{equation}
  \Phi^{2}E_{m} = \sqrt{|g|}g^{tt}g^{mn}\partial_{n}A_{t}\,,  
  \end{equation}
  where the right-hand side of the equation is computed using the fields of
  the complete spacetime solution.}

The total mass, electric charge and asymptotic value of the electrostatic
potential Eq.~(\ref{eq:electrostaticpotentialRNcase}) of these solutions
computed in the asymptotically-flat patch covered by these coordinates are

\begin{subequations}
  \begin{align}
    M
    & =
      \sum_{i=1}^{N}\left(\sigma_{i}+\kappa_{i}\right)\,,
    \\
    & \nonumber \\
    q
    & =
      2\sum_{i=1}^{N}\left(\sigma_{i}-\kappa_{i}\right)\,,
    \\
    & \nonumber \\
    P_{\infty}
    & =
      0\,.
  \end{align}
\end{subequations}

Observe that the total mass and charge satisfy the bound

\begin{equation}
M \geq |q|/2\,.
\end{equation}

Let us analyze the $\rho_{i}\equiv |\vec{x}-\vec{x}_{i}| \rightarrow 0$ limit
in which the metric seems to be singular. If either $\sigma_{i}$ or
$\kappa_{i}$ vanishes, the metric approaches the cylindrical metric
Eq.~(\ref{eq:cylindrical}) characteristic of extremal black holes. If neither
of them vanishes, the metric is singular in these coordinates and we must
analytically continue it using

\begin{equation}
  \frac{\rho_{i}}{(\sigma_{i}\kappa_{i})^{1/2}}
  =
\frac{(\sigma_{i}\kappa_{i})^{1/2}}{\rho_{i}'}\,.
\end{equation}

In the new coordinates that cover the $i$\textsuperscript{th}
asymptotically-flat region, the metric, the electric field and the
electrostatic potential (with the same choice of additive constant as in
Eq.~(\ref{eq:electrostaticpotentialRNcase})) take the form

\begin{subequations}
  \begin{align}
    ds^{2}_{\beta}
    & \sim
      \left(1 +\frac{\kappa_{i}A_{i}}{\rho_{i}'}\right)^{2}
      \left(1 +\frac{\sigma_{i}B_{i}}{\rho_{i}'}\right)^{2}
      \left(d\rho_{i}^{\prime\, 2}+\rho_{i}^{\prime\, 2}d\Omega_{(2)\, i}^{2}\right)\,,
    \\
    & \nonumber \\
    E_{m_{i}'}
    & \sim
      2\partial_{m_{i}'}\log{\left[\left(1+\frac{\kappa_{i}A_{i}}{\rho_{i}'}\right)
      \left(1 +\frac{\sigma_{i}B_{i}}{\rho_{i}'}\right)^{-1}\right]}\,,
    \\
    & \nonumber \\
    P
    & \sim
      2\log{\left[\left(1+\frac{\kappa_{i}A_{i}}{\rho_{i}'}\right)
      \left(1 +\frac{\sigma_{i}B_{i}}{\rho_{i}'}\right)^{-1}\right]}\,,
    \\
    & \nonumber \\
    A_{i}
    & =
      1+\sum_{j\neq i} \frac{\sigma_{i}}{|\vec{x}_{ji}|}\,,
    \\
    & \nonumber \\
    B_{i}
    & =
      1+\sum_{j\neq i} \frac{\kappa_{i}}{|\vec{x}_{ji}|}\,,
  \end{align}
\end{subequations}

\noindent
and we can compute the masses, electric charges and electrostatic potentials
at the $i^{th}$ asymptotic region\footnote{The sign of the charges computed in
  the $i^{th}$ asymptotic regions is the opposite if we take the orientation
  that woud be outward-pointing in that region. We have taken the sign that
  corresponds to the opposite orientation. Since the sign of the charge is
  purely conventional, only the relative signs of the charges at both sides of
  the Einstein--Rosend bridges are relevant. Our choice is such that the sum
  of the charges at both sides are equal. The opposite choice would give a
  vanishing ``total charge''.}

\begin{subequations}
  \begin{align}
    M_{i}
    & =
      \sigma_{i}+\kappa_{i}
      +\sum_{j\neq i}\frac{\sigma_{i}\kappa_{j}+\kappa_{i}\sigma_{j}}{|\vec{x}_{ij}|}\,,
    \\
    & \nonumber \\
    q_{i}
    & =
      2\left[\sigma_{i}-\kappa_{i}
      +\sum_{j\neq
      i}\frac{\sigma_{i}\kappa_{j}-\kappa_{i}\sigma_{j}}{|\vec{x}_{ij}|}\right]\,,
    \\
    & \nonumber \\
    P_{\infty_{i}}
    & =
      0\,,
  \end{align}
\end{subequations}

\noindent
and the bound

\begin{equation}
M_{i} \geq |q_{i}|/2\,,  
\end{equation}

\noindent
is also satisfied in the $i^{th}$ asymptotic region. It is easy to see that it
is saturated in the limit in which either $\kappa_{i}$ or $\sigma_{i}$
vanishes.

The total charge computed in the first asymptotic region, $q$, is equal to the
sum of these charges

\begin{equation}
  q
  =
  \sum_{i=1}^{N}q_{i}\,,  
\end{equation}

\noindent
because the electric field satisfies a Gauss law Eq.~(\ref{eq:Gauss1}).  The
energy does not satisfy a Gauss law and the interaction energy

\begin{equation}
  M_{int}
  =
  M-\sum_{i=1}^{N}M_{i}
  =
  -2\sum_{i\neq j}\frac{\sigma_{i}\kappa_{j}}{|\vec{x}_{ij}|}\,,
\end{equation}

\noindent
does not vanish in general, except for the single-center case.

We can solve for the integration constants for large separations
$|\vec{x}_{ij}|$ in terms of the physical parameters $M_{i},q_{i}$

\begin{subequations}
  \begin{align}
    \sigma_{i}
    & \sim
      \tfrac{1}{2}(M_{i}+q_{i}/2)\left[1
    -\tfrac{1}{2}\sum_{j\neq i}\frac{M_{j}-q_{j}/2}{|\vec{x}_{ij}|}
      +\mathcal{O}\left(1/|\vec{x}_{ij}|^{2}\right)\right]\,,
    \\
    & \nonumber \\
    \kappa_{i}
    & \sim
      \tfrac{1}{2}(M_{i}-q_{i}/2)\left[1
    -\tfrac{1}{2}\sum_{j\neq i}\frac{M_{j}+q_{j}/2}{|\vec{x}_{ij}|}
      +\mathcal{O}\left(1/|\vec{x}_{ij}|^{2}\right)\right]\,,
  \end{align}
\end{subequations}

\noindent
and express the interaction energy as

\begin{equation}
  M_{int}
  =
  -\sum_{i<j}\frac{M_{i}M_{j}}{|\vec{x}_{ij}|}
  +\sum_{i<j}\frac{q_{i}q_{j}/4}{|\vec{x}_{ij}|}
  +\mathcal{O}\left(1/|\vec{x}_{ij}|^{2}\right)\,,
\end{equation}

\noindent
\textit{i.e.}~as the sum of an attractive Newtonian contribution and a
repulsive (for like charges) Coulombian contribution. 

The total interaction energy vanishes when

\begin{equation}
  M_{i} = |q_{i}|/2\,,
  \hspace{.5cm}
  \forall i=1,\cdots, N\,,
\end{equation}

\noindent
which implies that either $\sigma_{i}=0$ or $\kappa_{i}=0$. As we have seen
above, in this case we cannot really define the individual masses $M_{i}$
because the region described by the new coordinates is not asymptotically
flat, but cylindrical. However, it is natural to accept that they approach the
extremal value $|q_{i}|/2$.

We would like to associate a non-vanishing interaction energy to a
non-equilibrium situation as in the previous examples. This is what the
results seem to indicate, but in this case we have to take into account that
the standard Komar charge is not closed on-shell and the mass does not satisfy
a Gauss law in stationary EM spacetimes. There is, however, an on-shell closed
\textit{generalized Komar charge}
\cite{Liberati:2015xcp,Ortin:2021ade,Mitsios:2021zrn} whose integral at
spatial infinity gives the combination $M-P_{\infty}q$. Since, in this case,
the electrostatic potential takes the same (vanishing) value in all the
spatial infinities and the electric charge always satisfy a Gauss law, it
follows that the mass must also satisfy a Gauss law and, when it does not
(\textit{i.e.}~when there is a non-vanishing interaction energy) the complete
spacetime cannot be stationary and must evolve towards a stationary spacetime
which should be a RN black hole.

In contrast with the Einstein-Dilaton case, single-center solutions always
have vanishing self-interaction energy, at least within the class described by
our ansatz. As we have mentioned, they are constant-time slices of the static
RN solutions.

\subsection{Magnetic and dyonic Reissner-Nordstr\"om  black holes}
\label{sec-magnetic-reissner-nordstrom}

Magnetic solutions can be obtained by simply using the electric-magnetic
duality transformations Eqs.~(\ref{eq:electric-magnetic}). In the particular
case of the Maxwell theory ($a=0$) those discrete transformations are a
particular case ($\alpha=\pi/2$) of the continuous group of electric-magnetic
duality transformations

\begin{equation}
    E'= \cos{\alpha}\,E +\sin{\alpha}\,B\,,
  \hspace{1cm}
  B'= -\sin{\alpha}\,E+\cos{\alpha}\,B\,,
\end{equation}

\noindent
which can be used to generate dyonic solutions out of the purely electric
ones.  These transformations leave the metric invariant and the solutions have
the same properties as the purely electric ones with the electric charges $q$
replaced by $\sqrt{q^{2}+p^{2}}$.

\subsection{Electric EMD black holes}
\label{sec-EMD}

In Ref.~\cite{Cvetic:2014vsa}, Cveti\v{c}, Gibbons and Pope found a solution
to Eqs.~(\ref{eq:Einstein}) and (\ref{eq:Gauss1}) with a non-trivial dilaton
field based on three harmonic functions, generalizing the solution found in
Ref.~\cite{Ortin:1995vg}.

Based on our experience with the RN solution, we could simply make an ansatz
based on a constant-time slice of the well-known electric, static,
spherically-symmetric, black-hole solutions of the EMD model (reviewed in
Appendix~\ref{eq:EMDBHs}) written in isotropic coordinates. These solutions
are given in Eqs.~(\ref{eq:electricEMDsolutionsisotropic}),
(\ref{eq:electricEMDsolutionsisotropicfunctions}) and
(\ref{eq:electricEMDsolutionsisotropicintegrationconstants}). The full metric
depends on 4 functions, but the spatial part, the vector field and the dilaton
field depend on just three: $\sigma,\kappa,\psi$. Furthermore, just as in the
RN case, the relations between the integration constants allow us to rewrite
the electric field in terms of just two, $\sigma$ and $\kappa$. Thus, we can
directly try the ansatz

\begin{subequations}
  \begin{align}
    \Phi^{4}
    & =
      (\sigma\kappa)^{\frac{2}{1+a^{2}}}\psi^{^{\frac{4a^{2}}{1+a^{2}}}}\,,
    \\
    & \nonumber \\
    E_{m}
    & =
      \frac{2}{\sqrt{1+a^{2}}}e^{a\phi/2}
      \partial_{m}\log{(\sigma/\kappa)}\,,
    \\
    & \nonumber \\
    e^{-a\phi}
    & =
    e^{-a\phi_{\infty}}\left(\frac{\sigma\kappa}{\psi^{2}}\right)^{\frac{2a^{2}}{1+a^{2}}}\,,
  \end{align}
\end{subequations}

\noindent
and we find that it satisfies Eqs.~(\ref{eq:Einstein}) and (\ref{eq:Gauss1})
when $\psi,\sigma,\kappa$ are arbitrary harmonic functions in 3-dimensional
Euclidean space $\mathbb{E}^{3}$

\begin{equation}
    \partial_{m}\partial_{m}(\psi,\sigma,\psi)
     =
      0\,,
\end{equation}

\noindent
where, as usual, we assume that the integration constants
$\psi_{i},\sigma_{i},\kappa_{i}$ are not negative.

These are the solutions of the EMD initial data problem found in
Ref.~\cite{Cvetic:2014vsa}.

Now we choose the harmonic functions

\begin{equation}
    \psi,\sigma,\kappa
     =
      1 +\sum_{i=1}^{N}\frac{\psi_{i},\sigma_{i},\kappa_{i}}{|\vec{x}-\vec{x}_{i}|}\,,
\end{equation}

\noindent
in the obvious notation.

The total mass, electric and scalar charges of these solutions computed in the
asymptotically-flat patch covered by these coordinates are\footnote{The
  electric charge can be defined in a coordinate-invariant way by
  \begin{equation}
    q = \frac{1}{4\pi}\int_{S^{2}}d^{2}\Sigma_{m}e^{-a\phi}E^{m}
    = -\rho^{2}\Phi^{2}e^{-a\phi}E_{\rho}\,. 
  \end{equation}
  Since this charge satisfies a Gauss law, this expression must be
  $\rho$-independent.

  There is no similar definition for the scalar charge, in general, and, as it
  is customary \cite{Gibbons:1996af}, we define it as the coefficient of
  $1/\rho$ in the asymptotic expansion of the dilaton.
}

\begin{subequations}
  \begin{align}
    M
    & =
      \frac{1}{1+a^{2}}\sum_{i=1}^{N}\left(\sigma_{i}+\kappa_{i}+2a^{2}\psi_{i}\right)\,,
    \\
    & \nonumber \\
    q
    & =
      \frac{2e^{-a\phi_{\infty}/2}}{\sqrt{1+a^{2}}}\sum_{i=1}^{N}\left(\sigma_{i}-\kappa_{i}\right)\,.
    \\
    & \nonumber \\
    \Sigma
    & =
       -\frac{2a}{1+a^{2}}\sum_{i=1}^{N}\left(\sigma_{i}+\kappa_{i}-2\psi_{i}\right)\,.
  \end{align}
\end{subequations}

Before studying the generic
$\rho_{i}\equiv |\vec{x}-\vec{x}_{i}|\rightarrow 0$ limit let us consider that
limit in the single center case, choosing $\vec{x}_{1}=0$ for
simplicity. Observe that this solution has 3 independent integration
constants, $\psi_{1},\sigma_{1},\kappa_{1}$, that describe 3 independent
physical constants (apart from $\phi_{\infty}$): $M,q,\Sigma$. Therefore,
these solutions have primary scalar hair described by $\Sigma$.

When the three integration constants are different from zero,
$\Phi^{4}\sim 1/\rho^{4}$ in that limit, and we can analytically extend the
metric using the coordinate $\rho'$ defined by

\begin{equation}
\frac{\rho'}{\Omega^{1/2}}
=
\frac{\Omega^{1/2}}{\rho}\,,
\hspace{1cm}
\Omega
\equiv
(\sigma_{1}\kappa_{1})^{\frac{1}{(1+a^{2})}}\psi_{1}^{\frac{2a^{2}}{1+a^{2}}}\,.
\end{equation}

In the patch described by the new coordinates, the solution takes the form

\begin{subequations}
  \label{eq:electricEMDsolutionsisotropicsinglecenterotherasymptoticregion}
  \begin{align}
    ds^{2}_{\beta}
    & =
      (\sigma'\kappa')^{\frac{2}{1+a^{2}}}\psi^{\prime\, \frac{4a^{2}}{1+a^{2}}}
      \left(d\rho^{\prime\, 2}+\rho^{\prime\, 2}d\Omega_{(2)}^{2}\right)\,,
    \\
    & \nonumber \\
    E_{m'}
    & =
      -\frac{2}{\sqrt{1+a^{2}}}e^{a\phi_{\infty}/2}e^{a\phi/2}
      \partial_{m'}\log{(\sigma'/\kappa')}\,,
    \\
    & \nonumber \\
    e^{-a\phi}
    & =
    e^{-a\phi_{\infty}}\left(\frac{\sigma_{1}\kappa_{1}}{\psi_{1}^{2}}\right)^{\frac{2a^{2}}{1+a^{2}}}\left(\frac{\sigma'\kappa'}{\psi^{\prime\,2}}\right)^{\frac{2a^{2}}{1+a^{2}}}\,,
  \end{align}
\end{subequations}

\noindent
where we have defined the functions

\begin{equation}
  \psi'
  =
  1+\frac{\Omega/\psi_{1}}{\rho'}\,,
  \hspace{1cm}
  \sigma'
  =
  1+\frac{\Omega/\sigma_{1}}{\rho'}\,,
  \hspace{1cm}
  \kappa'
  =
  1+\frac{\Omega/\kappa_{1}}{\rho'}\,.
\end{equation}

Thus, it is of the same form as in the other coordinate patch: asymptotically
flat and with mass, electric and magnetic charges and asymptotic value of the
dilaton given by

\begin{subequations}
  \begin{align}
    M'
    & =
      \frac{1}{1+a^{2}}\Omega\left(\frac{1}{\sigma_{1}}
      +\frac{1}{\kappa_{1}}
      +2a^{2}\frac{1}{\psi_{1}}\right)\,,
    \\
    & \nonumber \\
    q'
    & =
      q\,,
    \\
    & \nonumber \\
    \Sigma'
    & =
      -\frac{2a}{1+a^{2}}\Omega\left(
\frac{1}{\sigma_{1}}
      +\frac{1}{\kappa_{1}}
      -2\frac{1}{\psi_{1}}
\right)
    \\
    & \nonumber \\
    \phi_{\infty}'
    & =
      \phi_{\infty}
      -\frac{2a}{1+a^{2}}\log{\left(\frac{\sigma_{1}\kappa_{1}}{\psi_{1}^{2}}\right)}\,.
  \end{align}
\end{subequations}

As in the JNW case, the mass and scalar charge take different values in the
two asymptotic regions. The interaction energy $M_{int} = M-M'$ vanishes when

\begin{equation}
  \psi_{1}^{2}
  =
  \sigma_{1}\kappa_{1}\,,  
\end{equation}

\noindent
and the same condition makes the scalar charge (as well as the asymptotic
value of the dilaton) equal in both regions. When this condition holds, the
scalar charge is not independent anymore, but it is given in terms of the rest
of the physical constants by

\begin{subequations}
  \label{eq:sigmaversusMandq}
  \begin{align}
    \Sigma
    & \stackrel{a\neq 1}{=}
      -\frac{2a}{1-a^{2}}
      \left\{M
      -\sqrt{M^{2}-\tfrac{1}{4}(1-a^{2})e^{a\phi_{\infty}}q^{2}}
      \right\}\,,
      \\
    & \nonumber \\
    \Sigma
    & \stackrel{a=1}{=}
    -\frac{e^{\phi_{\infty}}q^{2}}{4M}\,.
  \end{align}
\end{subequations}

This is the value of the scalar charge in the complete, static, electric EMD
solutions reviewed in Appendix~\ref{eq:EMDBHs}. In stationary spacetimes one
can construct a definition of scalar charge which is coordinate-independent
and satisfies a Gauss law
\cite{Pacilio:2018gom,Ballesteros:2023iqb,Ballesteros:2023muf}. The value of
this charge for black holes with bifurcate horizons in EMD theories is
precisely the one we have found by demanding the vanishing of the interaction
energy. Furthermore, the fact that this value is the same in the two
asymptotic regions is in agreement with the Gauss law satisfied by this kind
of charge in stationary spacetimes. We conclude that, once more, vanishing
interaction energy points to the existence of a stationary solution of which
the initial data are a constant-time slice.

Notice that, in a given theory (\textit{i.e.}~for a given value of $a$) all
scalar charges have the same sign, just as all masses do. That sign is purely
conventional.

Let us now move to the $\rho_{i} \equiv |\vec{x}-\vec{x}_{i}| \rightarrow 0$
limit in the multi-center case. Defining the new coordinate

\begin{equation}
\frac{\rho_{i}'}{\Omega_{i}^{1/2}}
=
\frac{\Omega_{i}^{1/2}}{\rho_{i}}\,,
\hspace{1cm}
\Omega_{i}
\equiv
(\sigma_{i}\kappa_{i})^{\frac{1}{(1+a^{2})}}\psi_{i}^{\frac{2a^{2}}{1+a^{2}}}\,,
\end{equation}

\noindent
the solution takes the following form in that limit

\begin{subequations}
  \label{eq:electricEMDsolutionsisotropicotherasymptoticregion}
  \begin{align}
    ds^{2}_{\beta}
    & \sim
      (\sigma'\kappa')^{\frac{2}{1+a^{2}}}\psi^{\prime\, \frac{4a^{2}}{1+a^{2}}}
      \left(d\rho_{i}^{\prime\, 2}+\rho_{i}^{\prime\, 2}d\Omega_{(2)\, i}^{2}\right)\,,
    \\
    & \nonumber \\
    E_{m_{i}'}
    & \sim
      -\frac{2}{\sqrt{1+a^{2}}}e^{a\phi_{\infty}/2}e^{a\phi/2}
      \partial_{m_{i}'}\log{(\sigma'/\kappa')}\,,
    \\
    & \nonumber \\
    e^{-a\phi}
    & \sim
    e^{-a\phi_{\infty}}\left(\frac{\sigma_{i}\kappa_{i}}{\psi_{i}^{2}}\right)^{\frac{2a^{2}}{1+a^{2}}}\left(\frac{\sigma'\kappa'}{\psi^{\prime\,2}}\right)^{\frac{2a^{2}}{1+a^{2}}}\,,
  \end{align}
\end{subequations}

\noindent
where we have defined the functions

\begin{equation}
  \psi'
  =
  1+\frac{A_{i}\Omega_{i}/\psi_{1}}{\rho_{i}'}\,,
  \hspace{1cm}
  \sigma'
  =
  1+\frac{B_{i}\Omega_{i}/\sigma_{i}}{\rho_{i}'}\,,
  \hspace{1cm}
  \kappa'
  =
  1+\frac{C_{i}\Omega_{i}/\kappa_{i}}{\rho_{i}'}\,,
\end{equation}

\noindent
and the constants

\begin{equation}
  A_{i}
  =
  1 +\sum_{j\neq i}\frac{\psi_{j}}{|\vec{x}_{ij}|}\,,
  \hspace{1cm}
  B_{i}
  =
  1 +\sum_{j\neq i}\frac{\sigma_{j}}{|\vec{x}_{ij}|}\,,
  \hspace{1cm}
  C_{i}
  =
  1 +\sum_{j\neq i}\frac{\kappa_{j}}{|\vec{x}_{ij}|}\,.
  \hspace{1cm}
\end{equation}

The values of the physical constants computed in the new asymptotic region are

\begin{subequations}
  \begin{align}
    M_{i}
    & =
      \frac{1}{1+a^{2}}\Omega_{i}\left(
      \frac{B_{i}}{\sigma_{i}}
      +\frac{C_{i}}{\kappa_{i}}
      +2a^{2}\frac{A_{i}}{\psi_{i}}
      \right)\,,
    \\
    & \nonumber \\
    q_{i}
    & =
      -\frac{2e^{-a\phi_{\infty}/2}}{\sqrt{1+a^{2}}}\Omega_{i}
      \left(\frac{\sigma_{i}\kappa_{i}}{\psi_{i}^{2}}\right)^{\frac{a^{2}}{1+a^{2}}}
      \left(\frac{B_{i}}{\sigma_{i}}-\frac{C_{i}}{\kappa_{i}}\right)
      \nonumber \\
    & \nonumber \\
    & =
      -\frac{2e^{-a\phi_{\infty}/2}}{\sqrt{1+a^{2}}}
      \left(B_{i}\kappa_{i}-C_{i}\sigma_{i}\right)\,,
    \\
    & \nonumber \\
    \Sigma_{i}
    & =
      -\frac{2a}{1+a^{2}}\Omega_{i}\left(
      \frac{B_{i}}{\sigma_{i}}
      +\frac{C_{i}}{\kappa_{i}}
      -2\frac{A_{i}}{\psi_{i}}\,,
\right)
    \\
    & \nonumber \\
    \phi_{\infty\, i}'
    & =
      \phi_{\infty}
      -\frac{2a}{1+a^{2}}\log{\left(\frac{\sigma_{i}\kappa_{i}}{\psi_{i}^{2}}\right)}\,.
  \end{align}
\end{subequations}

It is not difficult to see that the total electric charge is conserved

\begin{equation}
  q
  =
  \sum_{i=1}^{N}q_{i}\,.
\end{equation}

\noindent
Neither the mass nor the scalar charge enjoy the same property. In particular,

\begin{equation}
  \begin{aligned}
    M_{int}
    & \equiv
    M-\sum_{i=1}^{N}M_{i}
    \\
    & \\
    & =
    \frac{1}{1+a^{2}}\sum_{i=1}^{N}\left[ \sigma_{i}+\kappa_{i}+2a^{2}\psi_{i}
      -\left(\frac{\sigma_{i}\kappa_{i}}{\psi_{i}^{2}}\right)^{-\frac{a^{2}}{1+a^{2}}}
      \left( B_{i}\kappa_{i} +C_{i}\sigma_{i}
        +2a^{2}A_{i}\psi_{i}\frac{\sigma_{i}\kappa_{i}}{\psi^{2}_{i}}\right) \right]\,.
  \end{aligned}
\end{equation}

Our study of the single-center case indicates that part of the interaction
energy can be understood as a property of each individual center and not to
the interaction with other centers, \textit{i.e.}~as a sort of
\textit{self-interaction energy}. This part of the interaction energy can be
removed by setting

\begin{equation}
  \psi_{i}^{2}
  =
  \sigma_{i}\kappa_{i}\,,
\end{equation}

\noindent
which, as can be easily checked, leads to the same asymptotic value for the
dilaton in all the asymptotic regions and also imposes the relations
Eqs.~(\ref{eq:sigmaversusMandq}) between the physical parameters of the
individual centers $\Sigma_{i},M_{i},q_{i},\phi_{\infty}$.

Then ,

\begin{equation}
    M_{int}
    =
    \frac{2}{1+a^{2}}\sum_{i\neq j}\frac{\sigma_{i}\kappa_{j} +2a^{2}(\sigma_{i}\kappa_{j}\sigma_{j}\kappa_{i})^{1/2}}{|\vec{x}_{ij}|}\,,
\end{equation}

\noindent
can be understood as the interaction energy between the non-extremal EMD black
holes described by the well-known solutions described in
Appendix~\ref{eq:EMDBHs}, which satisfy the relations
Eqs.~(\ref{eq:sigmaversusMandq}) and have no primary scalar hair.

In the $N=2$ case we can express the interaction energy in terms of the
physical constants of the individual centers as

\begin{equation}
    M_{int}
    =
-\frac{M_{1}M_{2}}{|\vec{x}_{12}|}  
+e^{a\phi_{\infty}}\frac{q_{1}q_{2}/4}{|\vec{x}_{12}|}  
-\frac{\Sigma_{1}\Sigma_{2}/4}{|\vec{x}_{12}|}
+\mathcal{O}(1/|\vec{x}_{12}|^{2})\,.
\end{equation}

This is one of our main results. Observe that the interaction between the
scalar charges of the centers is always attractive.

For identical centers with individual masses and charges $M,q,\Sigma$, the
no-force condition (zero interaction energy) takes the form

\begin{equation}
M^{2}                           %
-e^{a\phi_{\infty}}q^{2}/4+\Sigma^{2}/4
=
0\,.
\end{equation}

Since $\Sigma$ is the function of $M,q,\phi_{\infty}$ given in
Eqs.~(\ref{eq:sigmaversusMandq}), we can express the above condition as a
relation between $M,q,\phi_{\infty}$ only. This relation should take the form
of a BPS bound. We find, for all non-vanishing values of $a$, the following
relations for each of the centers:

\begin{equation}
  M
  =
  \frac{e^{a\phi_{\infty}/2}}{2\sqrt{1+a^{2}}}|q|\,,
\end{equation}

\noindent
so that

\begin{equation}
  \Sigma
  \stackrel{a\neq 1}{=}
  -4aM\,,
  \hspace{1cm}
  \Sigma
  \stackrel{a=1}{=}
  -2M\,.
\end{equation}

For these values of the physical parameters, the solutions are singular
($\psi_{1}=-M/2<0$) when $a\neq 1,0$ and the result must be understood as a
limit which, on the other hand, coincides with what is known of the extremal
EMD black holes. The solutions are regular for $a=0$ (ERN) and for $a=1$.

Finally, to end this section we notice that for particular values of the
parameters one obtains solutions which are cylindrical in the
$\rho\rightarrow 0$ limit so that there is no second asymptotically-flat
region. For a single center, this happens in the following cases:\footnote{In
  the cases in which $\Phi^{4}\sim \rho^{-3}$ or $\rho^{-1}$ in the
  $\rho\rightarrow 0$ limit, the analytical extension gives rise to an
  asymptotically-conical spacetime and we will not study them here.}

\begin{enumerate}
\item $\psi_{1}=0,\sigma_{1}\neq 0,\kappa_{1}\neq 0$, $a=\pm \sqrt{3}$.
\item $\psi_{1}\neq 0,\sigma_{1}= 0,\kappa_{1}\neq 0$, $a=0$, (and the
  $\sigma\leftrightarrow \kappa$ symmetric case).
\item $\psi_{1}\neq 0,\sigma_{1}= \kappa_{1} = 0$, $a=\pm 1$.
\end{enumerate}

\subsection{Magnetic and dyonic EMD  black holes}
\label{sec-magnetic-EMD}

As in the Reissner--Nordstr\"om case, we can transform a purely electric into
a purely magnetic solution using the discrete electric-magnetic duality
transformations Eqs.~(\ref{eq:electric-magnetic}). For $a\neq 0$, though,
these transformations are not part of a continuous electric-magnetic duality
group and one cannot generate dyonic from purely electric solutions with
them. The purely magnetic solutions one obtains from the purely electric ones
enjoy the same properties, upon the replacement of the electric charge by the
magnetic charge. In particular, the solutions corresponding to extremal black
holes are singular for all $a\neq 0,1$.

Dyonic black hole solutions have been found in fully analytic form for the
particular values $a=1$ \cite{Gibbons:1982ih,Gibbons:1987ps} and $a=\sqrt{3}$
\cite{Dobiasch:1981vh,Gibbons:1982ih,Gibbons:1985ac,Gibbons:1987ps,Gibbons:1994ff}. For
other values of $a$ it has been shown that black-hole solutions exist,
although with loss of analyticity in the horizon except for particular values
of $a$ \cite{Poletti:1995yq,Galtsov:2014wxl}.

In what follows, we are going to present and study new solutions to the
time-symmetric initial data problem for the case $a=1$. These are given in
terms of 4 harmonic functions in $\mathbb{E}^{3}$, $\sigma,\kappa,\psi,\tau$,

\begin{equation}
    \partial_{m}\partial_{m}(\psi,\sigma,\psi,\tau)
     =
      0\,,
\end{equation}

\noindent
by

\begin{subequations}
  \begin{align}
    \Phi^{4}
    & =
      \sigma\kappa\psi\tau
    \\
    & \nonumber \\
    E_{m}
    & =
      \alpha \sqrt{2} e^{\phi/2}
      \partial_{m}\log{(\sigma/\kappa)}\,,
    \\
    & \nonumber \\
    B_{m}
    & =
      \beta \sqrt{2} e^{-\phi/2}
      \partial_{m}\log{(\psi/\tau)}\,,
    \\
    & \nonumber \\
    e^{-\phi}
    & =
    e^{-\phi_{\infty}}\frac{\sigma\kappa}{\psi\tau}\,,
  \end{align}
\end{subequations}

\noindent
where $\alpha$ and $\beta$ square to one and indicate the sign of all the
electric and magnetic charges, respectively.

Asymptotically-flat solutions are obtained by choosing harmonic functions of
the form (in the obvious notation)

\begin{equation}
    \sigma,\kappa,\psi,\tau
     =
     1
     +\sum_{i=1}^{N}\frac{\sigma_{i},\kappa_{i},\psi_{i},\tau_{i}}{|\vec{x}-\vec{x}_{i}|}\,,
\end{equation}

\noindent
where all the coefficients  $\sigma_{i},\kappa_{i},\psi_{i},\tau_{i}$ are
assumed to be non-negative.

Setting $\tau=\psi$ we recover the purely electric $a=1$ solutions studied in
Section~\ref{sec-EMD} and with $\sigma\kappa=\psi\tau$ we recover the
Reissner--Nordstr\"om ones of Sections~\ref{sec-reissner-nordstrom}
and~\ref{sec-magnetic-reissner-nordstrom}.

\subsubsection{Single-center case}

In the $N=1$ case, the mass $M$, electric and magnetic charges $q,p$ and
scalar charge $\Sigma$ are given by\footnote{The
  magnetic charge can be defined in a coordinate-invariant way by
  \begin{equation}
    p = \frac{1}{4\pi}\int_{S^{2}}d^{2}\Sigma_{m}e^{\phi}B^{m}
    = -\rho^{2}\Phi^{2}e^{\phi}B_{\rho}\,. 
  \end{equation}
  Since this charge satisfies a Gauss law, this expression must be
  $\rho$-independent and the result of the integral should be the same for any
  2-sphere centered in the origin.}

\begin{subequations}
  \begin{align}
    M
    & =
      \tfrac{1}{2}\left(\sigma_{1}+\kappa_{1}+\psi_{1}+\tau_{1}\right)\,,
    \\
    & \nonumber \\
    q
    & =
      \alpha\sqrt{2} e^{-\phi_{\infty}/2}(\sigma_{1}-\kappa_{1})\,,
    \\
    & \nonumber \\
    p
    & =
      \beta\sqrt{2} e^{\phi_{\infty}/2}(\psi_{1}-\tau_{1})\,,
    \\
    & \nonumber \\
    \Sigma
    & =
      -\sigma_{1}-\kappa_{1}+\psi_{1}+\tau_{1}\,.
  \end{align}
\end{subequations}

Since the 4 integration constants $\sigma_{1},\kappa_{1},\psi_{1},\tau_{1}$
are independent, the above 4 charges, $M,q,p,\Sigma$ are also independent and
the solution has primary scalar hair.

When the 4 integration constants $\sigma_{1},\kappa_{1},\psi_{1},\tau_{1}$ are
all non-vanishing, there is another asymptotically-flat region in the
$\rho_{1}\equiv |\vec{x}-\vec{x}_{1}|\rightarrow 0$ limit. Using the same
methods we have employed in the previous cases, we find that the mass and
scalar charge of
the solution and the asymptotic value of the dilaton in that asymptotic region
are given by

\begin{subequations}
  \begin{align}
    M'
    & =
      \tfrac{1}{2}\Omega\left(\frac{1}{\sigma_{1}}+\frac{1}{\kappa_{1}}
      +\frac{1}{\psi_{1}}+\frac{1}{\tau_{1}}\right)\,,
    \\
    & \nonumber \\
    \Sigma'
    & =
      \Omega\left(-\frac{1}{\sigma_{1}}-\frac{1}{\kappa_{1}}
      +\frac{1}{\psi_{1}}+\frac{1}{\tau_{1}}\right)\,,
    \\
    & \nonumber \\
    \phi_{\infty}'
    & =
      \phi_{\infty}+\log{\left(\frac{\psi_{1}\tau_{1}}{\sigma_{1}\kappa_{1}}\right)}\,,
  \end{align}
\end{subequations}

\noindent
with

\begin{equation}
  \Omega^{2}
  \equiv
  \sigma_{1}\kappa_{1}\psi_{1}\tau_{1}\,,
\end{equation}

\noindent
while the electric and magnetic charges are equal because they satisfy Gauss laws.

The interaction mass $M_{int}=M-M'$ is generically different from zero. It
vanishes when the product of any two integration constants equals the product
of the other two. For the sake of concreteness we are going to choose

\begin{equation}
  \label{eq:nohairconstraint}
  \sigma_{1}\kappa_{1}
  =
  \psi_{1}\tau_{1}\,.
\end{equation}

It is not difficult to see that, as in all previous cases, when the
interaction energy vanishes, the scalar charge and the asymptotic value of the
scalar take the same values in both asymptotic regions $\Sigma=\Sigma'$,
$\phi_{\infty}'=\phi_{\infty}$. Furthermore, the above relation between the 4
integration constants implies the existence of a relation between the 4
charges that allows us to express the scalar charge in terms of the conserved
charges

\begin{equation}
  \label{eq:Sigmaasafunctioofcharges}
  \Sigma
  =
  \frac{e^{-\phi_{\infty}}p^{2}-e^{\phi_{\infty}}q^{2}}{4M}\,,
\end{equation}

\noindent
which indicates the absence of primary scalar charge. These solutions are
constant-time slices of the static solutions constructed in
Refs.~\cite{Gibbons:1982ih,Gibbons:1987ps}. When the constraint
Eq.~(\ref{eq:nohairconstraint}) is not imposed, the initial data describe a
dynamical black hole with primary scalar hair.

When only two of the integration constants are non-vanishing, the solution is
asymptotically cylindrical in the $\rho_{1}\rightarrow 0$ limit. Observe that,
in this case, there is always a (vanishing) product of two integration
constants which is equal to the (vanishing) product of the other two. This
leads to vanishing interaction energy and equal scalar charges and asymptotic
values of the scalar in both asymptotic regions. If we take, for instance,
$\kappa_{1}=\tau_{1}=0$, it is easy to see that the solutions are
constant-time slices of extremal, static, black holes saturating the bound

\begin{subequations}
  \begin{align}
    M
    & =
      \frac{e^{\phi_{\infty}/2}}{2\sqrt{2}}|q|
      +\frac{e^{-\phi_{\infty}/2}}{2\sqrt{2}}|p|\,,
    \\
    & \nonumber \\
    \Sigma
    & =
      -\frac{e^{\phi_{\infty}/2}}{2\sqrt{2}}|q|
      +\frac{e^{-\phi_{\infty}/2}}{2\sqrt{2}}|p|\,.    
  \end{align}
\end{subequations}

\subsubsection{The multi-center case}

For general values of $N$, we find the total charges

\begin{subequations}
  \begin{align}
    M
    & =
      \tfrac{1}{2}\sum_{i=1}^{N}\left(\sigma_{i}+\kappa_{i}+\psi_{i}+\tau_{i}\right)\,,
    \\
    & \nonumber \\
    q
    & =
      \alpha\sqrt{2} e^{-\phi_{\infty}/2}\sum_{i=1}^{N}(\sigma_{i}-\kappa_{i})\,,
    \\
    & \nonumber \\
    p
    & =
      \beta\sqrt{2} e^{\phi_{\infty}/2}\sum_{i=1}^{N}(\psi_{i}-\tau_{i})\,,
    \\
    & \nonumber \\
    \Sigma
    & =
    \sum_{i=1}^{N}\left(  -\sigma_{i}-\kappa_{i}+\psi_{i}+\tau_{i}\right)\,.
  \end{align}
\end{subequations}

When all the integration constants are different from zero there is another
asymptotically-flat region in each
$\rho_{i}\equiv |\vec{x}-\vec{x}_{i}| \rightarrow 0$ limit in which the
charges and the asymptotic values of the scalar take the values

\begin{subequations}
  \begin{align}
    M_{i}
    & =
      \tfrac{1}{2}
      \left(\sigma_{i}'+\kappa_{i}'+\psi_{i}'+\tau_{i}'\right)\,,
    \\
    & \nonumber \\
    q_{i}
    & =
      -\alpha\sqrt{2} e^{-\phi_{\infty\, i}/2}(\sigma_{i}'-\kappa_{i}')\,,
    \\
    & \nonumber \\
    p_{i}
    & =
      -\beta\sqrt{2} e^{\phi_{\infty\, i}/2}(\psi_{i}'-\tau_{i}')\,,
    \\
    & \nonumber \\
    \Sigma_{i}
    & =
    -\sigma_{i}'-\kappa_{i}'+\psi_{i}'+\tau_{i}',,
    \\
    & \nonumber \\
    \phi_{\infty\, i}
    & =
    \phi_{\infty}+\log{\left(\frac{\psi_{i}\tau_{i}}{\sigma_{i}\kappa_{i}}\right)}\,,
  \end{align}
\end{subequations}

\noindent
with

\begin{equation}
  \sigma_{i}'
  \equiv
  \Omega_{i}
  \left(1+\sum_{j\neq
      i}^{N}\frac{\sigma_{j}}{|\vec{x}_{ij}|}\right)/\sigma_{i}\,,
  \hspace{1cm}
  \Omega_{i}^{2}
  \equiv
  \sigma_{i}\kappa_{i}\psi_{i}\tau_{i}\,,
\end{equation}

\noindent
and similar expressions for $\kappa_{i}',\psi_{i}',\tau_{i}'$.

It is easy to check that the sum of the individual electric and magnetic
charges give the total electric and magnetic charge computed in the asymptotic
region that contains all the centers

\begin{equation}
  \sum_{i=1}^{N}q_{i}
  =
  q\,,
  \hspace{1cm}
  \sum_{i=1}^{N}p_{i}
  =
  p\,,
\end{equation}

\noindent
as they should. As usual, the same is not true for the individual masses
$M_{i}$ and the total mass $M$ nor for the scalar charges. In order to study
the interaction between the different black holes, we must remove the
self-interaction due to the presence of primary scalar hair, using the
condition Eq.~(\ref{eq:nohairconstraint}) for each center, \textit{i.e.},
imposing

\begin{equation}
  \label{eq:nohairconstraints}
  \sigma_{i}\kappa_{i}
  =
  \psi_{i}\tau_{i}\,,
  \,\,\,\,
  \forall i=1,\ldots,N\,.
\end{equation}

This condition makes all the asymptotic values of the scalar equal and,
furthermore, the scalar charge satisfies a Gauss law

\begin{equation}
  \sum_{i=1}^{N}\Sigma_{i}
  =
  \Sigma\,.
\end{equation}

The remaining interaction energy has the simple, exact, expression

\begin{equation}
  M_{int}
  =
  M-\sum_{i=1}^{N}M_{i}
  =
  -\sum_{i,j\neq i}^{N}\frac{\sigma_{j}\kappa_{i}+\kappa_{j}\sigma_{i}+\psi_{j}\tau_{i}+\tau_{j}\psi_{i}}{|\vec{x}_{ij}|}\,,
\end{equation}

\noindent
which can be rewritten in terms of the charges of two centers in the following
approximate way

\begin{equation}
  M_{int}
  =
  -\frac{M_{1}M_{2}}{|\vec{x}_{12}|}
  +\frac{e^{\phi_{\infty}}}{4}\frac{q_{1}q_{2}}{|\vec{x}_{12}|}
  +\frac{e^{-\phi_{\infty}}}{4}\frac{p_{1}p_{2}}{|\vec{x}_{12}|}
  -\frac{1}{4}\frac{\Sigma_{1}\Sigma_{2}}{|\vec{x}_{12}|}
  +\mathcal{O}\left(\frac{1}{|\vec{x}_{12}|^{2}}\right)\,,
\end{equation}

\noindent
which is commonly used in the literature for static, widely separated black
holes.

For identical black holes, the vanishing of the interaction energy leads to
the well-known BPS-type relation

\begin{equation}
  M^{2}-\frac{e^{\phi_{\infty}}q^{2}}{4}-\frac{e^{-\phi_{\infty}}p^{2}}{4}
  +\frac{1}{4}\Sigma^{2}
  =
  0\,.
\end{equation}

Since we have shown that the no-primary-hair condition implies
Eq.~(\ref{eq:Sigmaasafunctioofcharges}), it is not difficult to use this
relation to rewrite the above condition as

\begin{equation}
  \left[M
    -\tfrac{1}{2\sqrt{2}}\left(e^{\phi_{\infty}/2}|q|
      +e^{-\phi_{\infty}/2}|p|\right)\right]
  \left[M
    -\tfrac{1}{2\sqrt{2}}\left(e^{\phi_{\infty}/2}|q|
      -e^{-\phi_{\infty}/2}|p|\right)\right]
  =
  0\,,
\end{equation}

\noindent
which shows that the condition is satisfied when the mass equals either of the
two central charge skew eigenvalues\footnote{These combinations have that
  interpretation when the $a=1$ theory is embedded in pure $\mathcal{N}=4,d=4$
  ungauged supergravity. See
  Refs.~\cite{Kallosh:1992ii,Kallosh:1993yg,Bergshoeff:1996gg,Lozano-Tellechea:1999lwm}.}

\begin{equation}
  \mathcal{Z}_{\pm}
  \equiv
  \tfrac{1}{2\sqrt{2}}\left(e^{\phi_{\infty}/2}|q|
      \pm e^{-\phi_{\infty}/2}|p|\right)\,.
\end{equation}

It is not difficult to see that, in these cases, at least two integration
coefficients of the center must vanish and that in the $\rho_{i}\rightarrow 0$
limit the solution becomes cylindrical, corresponding to the infinite throat
of an extremal black hole.

 \section{Discussion}
\label{sec-discussion}

In this paper we have shown how our interpretation of the non-vanishing
\textit{self-interaction energy} is related to the existence of primary scalar
hair and to the necessary evolution of the black hole solution. We would like
to stress that the black-hole initial data with primary scalar hair are
completely regular. However this does not contradict the no-hair conjecture,
which only refers to the endpoint of gravitational collapse and therefore, to
the stationary state which will be eventually reached.

This raises some important questions that we have already posed in the
previous sections and which fall out of the scope of this paper: will the
final state of the evolution of the black holes with primary hair contain some
primary hair or will it be completely lost (radiated away)? Will the final
state be a regular black hole?

Studying the evolution of the initial data describing single black holes with
primary hair which we have provided should be possible using numerical
methods. We believe there is a great deal of black-hole physics to be learned
from this problem.

\section*{Acknowledgments}

UKBV would like to thank \'Alvaro de la Cruz--Dombriz for his support and
acknowledge the hospitality of the IFT-UAM/CSIC in Madrid during the earliest
stages in the preparation of the manuscript. TOM would also like to thank
\'Alvaro de la Cruz--Dombriz for bringing him the opportunity to participate
in the Erasums+ program.  UKBV acknowledges financial support from the
National Research Foundation of South Africa, Grant number PMDS22063029733,
from the University of Cape Town Postgraduate Funding Office, the University
of Groningen and the Erasmus+ KA107 International Credit Mobility
Programme. This work has been supported in part by the MCI, AEI, FEDER (UE)
grants PID2021-125700NB-C21 (``Gravity, Supergravity and Superstrings''
(GRASS)), and IFT Centro de Excelencia Severo Ochoa CEX2020-001007-S.  TO
wishes to thank M.M.~Fern\'andez for her permanent support.

\appendix

\section{Static, electric, black-hole solutions of the EMD model}
\label{eq:EMDBHs}

The static, purely electric, black-hole solutions\footnote{Solutions including
  primary scalar hair were found in Ref.~\cite{Agnese:1994zx}. They are
  singular when the primery hair is present.}  of the EMD model presented in
Eq.~(\ref{eq:actiondilatonBHs})  were found in
Refs.~\cite{Gibbons:1982ih,Gibbons:1984kp,Holzhey:1991bx} and can be written
in the convenient form

\begin{subequations}
  \label{eq:electricEMDsolutions}
  \begin{align}
    ds^{2}
    & =
    H^{-\frac{2}{1+a^{2}}}W dt^{2}
    -H^{\frac{2}{1+a^{2}}}\!\left[W^{-1}dr^{2} +r^{2}d\Omega_{(2)}^{2}
    \right]\,,
    \\
    & \nonumber \\
    A_{t}
    & =
    \alpha e^{a\phi_{\infty}/2}(H^{-1}-1)\,,
    \\
    & \nonumber \\
    e^{-\phi}
    & =
    e^{-\phi_{\infty}} H^{\frac{2a}{1+a^{2}}}\,,
  \end{align}
\end{subequations}

\noindent
where the functions $H$ and $W$ (often called the \textit{blackening factor})
take the form

\begin{equation}
  H
  =
  1+\frac{h}{r}\,,
  \hspace{1.5cm}
  W
  =
  1+\frac{\omega}{r}\,,
\end{equation}

\noindent
and where the integration constants $h,\omega,\alpha$ are related by

\begin{equation}
  \label{eq:omegahalpha}
  \omega
  =
  h\!\left[1 -(1+a^{2})(\alpha/2)^{2} \right]\,.  
\end{equation}

The physical parameters of these solutions are the ADM mass $M$, the electric
charge $q$ and the asymptotic value of the scalar (\textit{modulus})
$\phi_{\infty}$. The integration constants are given in terms of them by

\begin{equation}
  \begin{aligned}
    h
    & =
    -\frac{a^{2}+1}{a^{2}-1}\left\{M
      -\sqrt{M^{2}+\tfrac{1}{4}(a^{2}-1)e^{a\phi_{\infty}}q^{2}}\right\}\,,
    \\
    & \\
    \omega
    & =
-\frac{2}{a^{2}-1}\left\{a^{2}M
      -\sqrt{M^{2}+\tfrac{1}{4}(a^{2}-1)e^{a\phi_{\infty}}q^{2}}\right\}\,,
    \\
    & \\
    \alpha
    & =
-4qe^{a\phi_{\infty}/2}/h\,,
  \end{aligned}
\end{equation}

\noindent
for $a\neq 1$ and 

\begin{equation}
  \begin{aligned}
    h
    & =
    \frac{4e^{\phi_{\infty}}q^{2}}{M}\,,
    \\
    & \\
    \omega
    & =
    -2\frac{M^{2}-2e^{\phi_{\infty}}q^{2}}{M}\,,
    \\
    & \\
    \alpha
    & =
    -e^{-\phi_{\infty}/2}M/q\,,
  \end{aligned}
\end{equation}

\noindent
for $a=1$.

The scalar charge $\Sigma$, conventionally defined by

\begin{equation}
  \label{eq:Sigmadefined}
  \phi
  \sim
  \phi_{\infty}+\frac{\Sigma}{r}\,,  
\end{equation}

\noindent
takes the value

\begin{equation}
  \Sigma
  =
  -\frac{2ah}{a^{2}+1}\,,
\end{equation}

\noindent
and we can write the integration constant $\omega$ in terms of the physical
constants $M$ and $\Sigma$ as

\begin{equation}
  \omega  =   -\frac{1}{a}\left[2aM -\Sigma\right]\,.
\end{equation}

We have chosen the sign of the square roots in $h$ and $\omega$ so that $h$ is
always positive and $\omega$ is always negative when the non-extremality
condition

\begin{equation}
  M^{2}
  >
  \frac{4}{a^{2}+1}e^{a\phi_{\infty}}q^{2}\,,
\end{equation}

\noindent
is met. In that case, there is an event horizon at

\begin{equation}
  r
  =
  -\omega\equiv r_{0}\,.  
\end{equation}

\noindent
Its Bekenstein-Hawking entropy and Hawking temperature are given by 

\begin{eqnarray}
  S
  & = &
        \pi r_{0}^{\frac{2a^{2}}{a^{2}+1}}(r_{0}+h)^{\frac{2}{a^{2}+1}}\,,
  \\
  & & \nonumber \\
        T
  & = &
        \frac{r_{0}}{4S}\,.
\end{eqnarray}

In the extremal limit ($\omega=r_{0}=0$) all the solutions
become singular except for $a=0$.

\section{Isotropic coordinates}

Often, we find 3-dimensional metrics conformal to one of the form

\begin{equation}
  W^{-1}dr^{2} +r^{2}d\Omega^{2}_{(2)}\,,
  \hspace{1cm}
  W
  =
  \left(1+\frac{\omega}{r}\right)\,.
\end{equation}

The coordinate transformation

\begin{equation}
  \label{eq:isotropictrans}
  r
  =
  \rho\left(1-\frac{\omega/4}{\rho}\right)^{2}\,,
\end{equation}

\noindent
brings it to the form

\begin{equation}
  \left(1-\frac{\omega/4}{\rho}\right)^{4}
  \left(d\rho^{2} +\rho^{2}d\Omega^{2}_{(2)}\right)\,.  
\end{equation}

\subsection{The Schwarzschild solution in isotropic coordinates}

Making this coordinate transformation in the Schwarzschild metric

\begin{equation}
  ds^{2}
  =
  \left(1-\frac{2M}{r}\right)dt^{2} -\left(1-\frac{2M}{r}\right)^{-1}dr^{2}
  -r^{2}d\Omega^{2}_{(2)}\,,
\end{equation}

\noindent
($\omega = -2M$), it takes the form

\begin{equation}
  \label{eq:Schwarzschildisotropic}
  ds^{2}
  =
\frac{\left(1-\frac{M/2}{\rho}\right)^{2}}{\left(1+\frac{M/2}{\rho}\right)^{2}}
dt^{2}
-\left(1+\frac{M/2}{\rho}\right)^{4}
  \left(d\rho^{2} +\rho^{2}d\Omega^{2}_{(2)}\right)\,.
\end{equation}

\subsection{The Janis-Newman-Winicour solution in isotropic coordinates}

Upon the transformation Eq.~(\ref{eq:isotropictrans}), the
Janis--Newman--Winicour (JNW) solution \cite{Janis:1968zz}

\begin{subequations}
  \label{eq:JNW}
  \begin{align}
  ds^{2}
  & =
  W^{-\alpha}dt^{2} - W^{1+\alpha}\left[W^{-1}dr^{2}
    +r^{2}d\Omega^{2}_{(2)}\right]\,,
    \\
    & \nonumber \\
    \phi
    & =
      \phi_{0} \pm \sqrt{1-\alpha^{2}}\log{W}\,,
  \end{align}
\end{subequations}

\noindent
takes the form

\begin{subequations}
  \label{eq:JNWisotropic}
  \begin{align}
  ds^{2}
  & =
    (\psi/\chi)^{-2\alpha}dt^{2}
    -(\psi\chi)^{2} (\psi/\chi)^{2\alpha}
    \left(d\rho^{2} +\rho^{2}d\Omega^{2}_{(2)}\right)\,,
    \\
    & \nonumber \\
    \phi
    & =
      \phi_{0} \pm 2\sqrt{1-\alpha^{2}}\log{(\psi/\chi)}\,,
  \end{align}
\end{subequations}

\noindent
with

\begin{equation}
  \psi
  =
  1+\frac{\omega/4}{\rho}\,,
  \hspace{1cm}
  \chi
  =
  1-\frac{\omega/4}{\rho}\,.
\end{equation}

\subsection{The electric Einstein--Maxwell--Dilaton solution in isotropic
  coordinates}
\label{sec-EMDisotropic}

Performing the same transformation in the electric EMD solutions in
Eqs.~(\ref{eq:electricEMDsolutions}) they are brought to the form

\begin{subequations}
  \label{eq:electricEMDsolutionsisotropic}
  \begin{align}
    ds^{2}
    & =
    (\sigma\kappa)^{-\frac{2}{1+a^{2}}}\psi^{-2+\frac{4}{1+a^{2}}}\chi^{2} dt^{2}
      -(\sigma\kappa)^{\frac{2}{1+a^{2}}}\psi^{^{\frac{4a^{2}}{1+a^{2}}}}\!
      \left(d\rho^{2} +\rho^{2}d\Omega^{2}_{(2)}\right)\,,
    \\
    & \nonumber \\
    A_{t}
    & =
    \alpha e^{a\phi_{\infty}/2}\left(\frac{\psi^{2}}{\sigma\kappa}-1\right)\,,
    \\
    & \nonumber \\
    \phi
    & =
    \phi_{\infty} -\frac{2a}{1+a^{2}}\log{\left(\frac{\sigma\kappa}{\psi^{2}}\right)}\,,
  \end{align}
\end{subequations}

\noindent
with

\begin{equation}
  \label{eq:electricEMDsolutionsisotropicfunctions}
    \psi
    =
    1+\frac{\psi_{1}}{\rho}\,,
    \hspace{.8cm}
    \chi
     =
      1+\frac{\chi_{1}}{\rho}\,,
    \hspace{.8cm}
    \sigma
    =
      1+\frac{\sigma_{1}}{\rho}\,,
      \hspace{.8cm}
    \kappa
    =
      1+\frac{\kappa_{1}}{\rho}\,,
    \end{equation}

\noindent
and

\begin{equation}
  \label{eq:electricEMDsolutionsisotropicintegrationconstants}
  \begin{aligned}
    \psi_{1} & = -\omega/4\,, \hspace{.5cm} & \chi_{1} & =\omega/4\,,
    \\
    & & & \\
    \sigma_{1} & = \frac{\omega^{2}}{4h\left( 1+\alpha\sqrt{1+a^{2}}/2\right)^{2}}\,,
    \hspace{.5cm}
    & \kappa_{1} & =
    \frac{\omega^{2}}{4h\left( 1-\alpha\sqrt{1+a^{2}}/2\right)^{2}}\,,
  \end{aligned}
\end{equation}

\noindent
where the constants $\omega,h,\alpha$ are related by the constraint
Eq.~(\ref{eq:omegahalpha}).

Observe that

\begin{equation}
  \psi_{1}
  =
  -\chi_{1}\,,
  \hspace{1cm}
  \frac{\psi_{1}^{2}}{\sigma_{1}\kappa_{1}} = 1\,.  
\end{equation}

These solutions include the Reissner-Nordstr\"om one for $a=0$. It takes the form

\begin{subequations}
  \label{eq:electricRNsolutionsisotropic}
  \begin{align}
    ds^{2}
    & =
    \left(\frac{\psi\chi}{\sigma\kappa}\right)^{2} dt^{2}
      -(\sigma\kappa)^{2}\!\left(d\rho^{2} +\rho^{2}d\Omega^{2}_{(2)}\right)\,,
    \\
    & \nonumber \\
    A_{t}
    & =
    \alpha \left(\frac{\psi^{2}}{\sigma\kappa}-1\right)\,,
  \end{align}
\end{subequations}

\noindent
with

\begin{equation}
  \label{eq:electricRNsolutionsisotropicintegrationconstants}
  \begin{aligned}
    \psi_{1} & = -\omega/4\,, \hspace{.5cm} & \chi_{1} & =\omega/4\,,
    \\
    & & & \\
    \sigma_{1} & = \frac{\omega^{2}}{4h\left( 1+\alpha/2\right)^{2}}\,,
    \hspace{.5cm}
    & \kappa_{1} & =
    \frac{\omega^{2}}{4h\left( 1-\alpha/2\right)^{2}}\,.
  \end{aligned}
\end{equation}

For $\alpha=0$ one recovers the Schwarzschild solution which has
$\sigma=\kappa$.

This famiy of solutions does not include the JNW one. A more general solution
such as the Agnese-La Camera solution \cite{Agnese:1994zx} is needed to cover
all the possiblities.



\begin{thebibliography}{99}

\bibitem{Einstein:1927}
A.~Einstein and J.~Grommer,
``Allgemeine Relativit\"atstheorie und Bewegungsgesetz,'' 
S.\ B.\ preuss.\ Akad.\ Wiss.\ \textbf{1} (1927).

\bibitem{Einstein:1938yz}
A.~Einstein, L.~Infeld and B.~Hoffmann,
``The Gravitational equations and the problem of motion,''
Annals Math.\  {\bf 39} (1938) 65.
\doi{10.2307/1968714}

\bibitem{Fock:1939}
V.~A.~Fock,
``Sur le mouvement de masses finies d'apres la th\'eorie de 
gravitation einsteinienne,''
J.\ Phys.\ U.S.S.R. {\bf 1} (1939) 81-116.

\bibitem{Einstein:1940mt}
A.~Einstein and L.~Infeld,
``The Gravitational equations and the problem of motion II,''
Annals Math.\ {\bf 41} (1940) 455.
\doi{10.2307/1969015}

\bibitem{Einstein:1949}
A.~Einstein and L.~Infeld,
``On the motion of particles in general relativity theory,''
Can.\ J.\ Math.\ {\bf 1} (1949) 209-241.

\bibitem{Papapetrou:1951}
A.~Papapetrou,
``Equations of motion in General Relativity,''
Proc.\ Phys.\ Soc.\, A {\bf 64} (1951) 57-75.

\bibitem{LIGOScientific:2016aoc}
B.~P.~Abbott \textit{et al.} [LIGO Scientific and Virgo],
``Observation of Gravitational Waves from a Binary Black Hole Merger,''
Phys. Rev. Lett. \textbf{116} (2016) no.6, 061102
\doi{10.1103/PhysRevLett.116.061102}
[\arxiv{1602.03837} [gr-qc]].

\bibitem{Brill:1959zz}
D.~R.~Brill,
``On the positive definite mass of the Bondi-Weber-Wheeler
time-symmetric gravitational waves,''
Annals Phys. \textbf{7} (1959), 466-483
\doi{10.1016/0003-4916(59)90055-7}

\bibitem{Brill:1963yv}
D.~R.~Brill and R.~W.~Lindquist,
``Interaction energy in geometrostatics,''
Phys. Rev. \textbf{131} (1963), 471-476
\doi{10.1103/PhysRev.131.471}

\bibitem{Gibbons:1972ym}
G.~W.~Gibbons,
``The time symmetric initial value problem for black holes,''
Commun. Math. Phys. \textbf{27} (1972), 87-102
\doi{10.1007/BF01645614}

\bibitem{Ortin:1995vg}
T.~Ort\'{\i}n,
``Time symmetric initial data sets in 4-d dilaton gravity,''
Phys. Rev. D \textbf{52} (1995), 3392-3405
\doi{10.1103/PhysRevD.52.3392}
[\hepth{9501094} [hep-th]].

\bibitem{Cvetic:2014vsa}
M.~Cveti\v{c}, G.~W.~Gibbons and C.~N.~Pope,
``Super-Geometrodynamics,''
JHEP \textbf{03} (2015), 029
\doi{10.1007/JHEP03(2015)029}
[\arxiv{1411.1084} [hep-th]].

\bibitem{Cremonini:2023vwf}
S.~Cremonini, M.~Cveti\v{c}, C.~N.~Pope and A.~Saha,
``Mass and force relations for Einstein-Maxwell-dilaton black holes,''
Phys. Rev. D \textbf{107} (2023) no.12, 126023
\doi{10.1103/PhysRevD.107.126023}
[\arxiv{2304.04791} [hep-th]].

\bibitem{Einstein:1935tc}
A.~Einstein and N.~Rosen,
``The Particle Problem in the General Theory of Relativity,''
Phys. Rev. \textbf{48} (1935), 73-77
\doi{10.1103/PhysRev.48.73}

\bibitem{Komar:1958wp}
A.~Komar,
``Covariant conservation laws in general relativity,''
Phys. Rev. \textbf{113} (1959), 934-936
\doi{10.1103/PhysRev.113.934}

\bibitem{Liberati:2015xcp}
S.~Liberati and C.~Pacilio,
``Smarr Formula for Lovelock Black Holes: a Lagrangian approach,''
Phys. Rev. D \textbf{93} (2016) no.8, 084044
\doi{10.1103/PhysRevD.93.084044}
[\arxiv{1511.05446} [gr-qc]].

\bibitem{Ortin:2021ade}
T.~Ort\'{\i}n,
``Komar integrals for theories of higher order
in the Riemann curvature and black-hole chemistry,''
JHEP \textbf{08} (2021), 023
\doi{10.1007/JHEP08(2021)023}
[\arxiv{2104.10717} [gr-qc]].

\bibitem{Mitsios:2021zrn}
D.~Mitsios, T.~Ort\'{\i}n and D.~Pere\~niguez,
``Komar integral and Smarr formula for axion-dilaton black
holes versus S duality,''
JHEP \textbf{08} (2021), 019
\doi{10.1007/JHEP08(2021)019}
[\arxiv{2106.07495} [hep-th]].

\bibitem{Pacilio:2018gom}
C.~Pacilio,
``Scalar charge of black holes in Einstein-Maxwell-dilaton theory,''
Phys. Rev. D \textbf{98} (2018) no.6, 064055
\doi{10.1103/PhysRevD.98.064055}
[\arxiv{1806.10238} [gr-qc]].

\bibitem{Ballesteros:2023iqb}
R.~Ballesteros, C.~G\'omez-Fayr\'en, T.~Ort\'{\i}n and M.~Zatti,
``On scalar charges and black hole thermodynamics,''
JHEP \textbf{05} (2023), 158
\doi{10.1007/JHEP05(2023)158}
[\arxiv{2302.11630} [hep-th]].

\bibitem{Coleman:1991ku}
S.~R.~Coleman, J.~Preskill and F.~Wilczek,
``Quantum hair on black holes,''
Nucl. Phys. B \textbf{378} (1992), 175-246
\doi{10.1016/0550-3213(92)90008-Y}
[\hepth{9201059} [hep-th]].

\bibitem{Janis:1968zz}
A.~I.~Janis, E.~T.~Newman and J.~Winicour,
``Reality of the Schwarzschild Singularity,''
Phys. Rev. Lett. \textbf{20} (1968), 878-880
\doi{10.1103/PhysRevLett.20.878}

\bibitem{Gibbons:1982ih}
G.~W.~Gibbons,
``Antigravitating Black Hole Solitons with Scalar Hair in N=4 Supergravity,''
Nucl. Phys. B \textbf{207} (1982), 337-349
\doi{10.1016/0550-3213(82)90170-5}

\bibitem{Gibbons:1984kp}
  G.~W.~Gibbons,
``Aspects of Supergravity Theories,''
 (three lectures)  in {\sl Supersymmetry, Supergravity
   and Related Topics}, Eds. F. del \'{A}guila, J. de Azc\'{a}rraga
 and L. Ib\'{a}\~nez, Singapore:
World Scientific (1985), p. 147.

\bibitem{Holzhey:1991bx}
C.~F.~E.~Holzhey and F.~Wilczek,
``Black holes as elementary particles,''
Nucl. Phys. B \textbf{380} (1992), 447-477
\doi{10.1016/0550-3213(92)90254-9}
[\hepth{9202014} [hep-th]].

\bibitem{Misner:1957mt}
C.~W.~Misner and J.~A.~Wheeler,
``Classical physics as geometry: Gravitation, electromagnetism,
unquantized charge, and mass as properties of curved empty space,''
Annals Phys. \textbf{2} (1957), 525-603
\doi{10.1016/0003-4916(57)90049-0}

\bibitem{Gibbons:1996af}
G.~W.~Gibbons, R.~Kallosh and B.~Kol,
``Moduli, scalar charges, and the first law of black hole thermodynamics,''
Phys. Rev. Lett. \textbf{77} (1996), 4992-4995
\doi{10.1103/PhysRevLett.77.4992}
[\hepth{9607108} [hep-th]].

\bibitem{Ballesteros:2023muf}
R.~Ballesteros and T.~Ort\'\i{}n,
``Hairy black holes, scalar charges and extended thermodynamics,''
Class. Quant. Grav. \textbf{41} (2024) no.5, 055007
\doi{10.1088/1361-6382/ad210a}
[\arxiv{2308.04994} [gr-qc]].

\bibitem{Gibbons:1987ps}
G.~W.~Gibbons and K.~i.~Maeda,
``Black Holes and Membranes in Higher Dimensional
Theories with Dilaton Fields,''
Nucl. Phys. B \textbf{298} (1988), 741-775
\doi{10.1016/0550-3213(88)90006-5}

\bibitem{Dobiasch:1981vh}
P.~Dobiasch and D.~Maison,
``Stationary, Spherically Symmetric Solutions of Jordan's Unified Theory of Gravity and Electromagnetism,''
Gen. Rel. Grav. \textbf{14} (1982), 231-242
\doi{10.1007/BF00756059}

\bibitem{Gibbons:1985ac}
G.~W.~Gibbons and D.~L.~Wiltshire,
``Black Holes in Kaluza-Klein Theory,''
Annals Phys. \textbf{167} (1986), 201-223
[erratum: Annals Phys. \textbf{176} (1987), 393]
\doi{10.1016/S0003-4916(86)80012-4}

\bibitem{Gibbons:1994ff}
G.~W.~Gibbons and R.~E.~Kallosh,
``Topology, entropy and Witten index of dilaton black holes,''
Phys. Rev. D \textbf{51} (1995), 2839-2862
\doi{10.1103/PhysRevD.51.2839}
[\hepth{9407118} [hep-th]].


\bibitem{Poletti:1995yq}
S.~J.~Poletti, J.~Twamley and D.~L.~Wiltshire,
``Dyonic Dilaton black holes,''
Class. Quant. Grav. \textbf{12} (1995), 1753-1770
[erratum: Class. Quant. Grav. \textbf{12} (1995), 2355]
\doi{10.1088/0264-9381/12/7/017}
[\hepth{9502054} [hep-th]].

\bibitem{Galtsov:2014wxl}
D.~Gal'tsov, M.~Khramtsov and D.~Orlov,
``\textquotedblleft{}Triangular\textquotedblright{}
extremal dilatonic dyons,''
Phys. Lett. B \textbf{743} (2015), 87-92
\doi{10.1016/j.physletb.2015.02.017}
[\arxiv{1412.7709} [hep-th]].

\bibitem{Kallosh:1992ii}
R.~Kallosh, A.~D.~Linde, T.~Ort\'{\i}n, A.~W.~Peet and A.~Van Proeyen,
``Supersymmetry as a cosmic censor,''
Phys. Rev. D \textbf{46} (1992), 5278-5302
\doi{10.1103/PhysRevD.46.5278}
[\hepth{9205027} [hep-th]].

\bibitem{Kallosh:1993yg}
R.~Kallosh and T.~Ort\'{\i}n,
``Charge quantization of axion - dilaton black holes,''
Phys. Rev. D \textbf{48} (1993), 742-747
\doi{10.1103/PhysRevD.48.742}
[\hepth{9302109} [hep-th]].

\bibitem{Bergshoeff:1996gg}
E.~Bergshoeff, R.~Kallosh and T.~Ort\'{\i}n,
``Stationary axion / dilaton solutions and supersymmetry,''
Nucl. Phys. B \textbf{478} (1996), 156-180
\doi{10.1016/0550-3213(96)00408-7}
[\hepth{9605059} [hep-th]].

\bibitem{Lozano-Tellechea:1999lwm}
E.~Lozano-Tellechea and T.~Ort\'{\i}n,
``The General, duality invariant family of
nonBPS black hole solutions of N=4, D = 4 supergravity,''
Nucl. Phys. B \textbf{569} (2000), 435-450
\doi{10.1016/S0550-3213(99)00762-2}
[\hepth{9910020} [hep-th]].

\bibitem{Agnese:1994zx}
A.~G.~Agnese and M.~La Camera,
``General spherically symmetric solutions in charged dilaton gravity,''
Phys. Rev. D \textbf{49} (1994), 2126-2128
\doi{10.1103/PhysRevD.49.2126}










\end{thebibliography}
\end{document}